\documentclass[figure]{pasj00}
 \draft

\begin{document}
\SetRunningHead{Shimajiri et al. 2010}{Survey for Possible External Triggers of Star Formation}

\title{New Panoramic View of $^{12}$CO and 1.1 mm Continuum Emission in the Orion A Molecular Cloud. I.  \\
Survey Overview and Possible External Triggers of Star Formation}

\author{Yoshito Shimajiri$^{1,2}$, 
Ryohei Kawabe$^{2,3}$, 
Shigehisa Takakuwa$^{4}$,
Masao Saito$^{3,5}$, 
Takashi Tsukagoshi$^{6}$, 
Munetake Momose$^{7}$, 
Norio Ikeda$^{8}$,  
E. Akiyama $^{7}$, 
Austermann, J. E. $^{9,10}$
H. Ezawa$^{3}$, 
K. Fukue$^{1,5}$, 
M. Hiramatsu$^{4}$, 
D. Hughes$^{11}$, 
Y. Kitamura$^{8}$, 
K. Kohno$^{6}$, 
Y. Kurono$^{3,5}$, 
Scott, K. S.$^{12}$,
G. Wilson$^{9}$, 
A. Yoshida$^{13}$,
M.S. Yun$^{9}$,  \\

\thanks{Last update: Sep,  20, 2010}}
\affil{$^{1}$ Department of Astronomy, School of Science, University of Tokyo, Bunkyo, Tokyo 113-0033, Japan}

\affil{$^{2}$ Nobeyama Radio Observatory, Minamimaki, Minamisaku, Nagano 384-1805, Japan}

\affil{$^{3}$ National Astronomical Observatory of Japan, Osawa 2-21-1, Mitaka, Tokyo 181-8588, Japan}

\affil{$^{4}$ Academia Sinica Institute of Astronomy and Astrophysics, P.O. Box 23-141, Taipei 10617, Taiwan}

\affil{$^{5}$ ALMA Project Office, National Astronomical Observatory of Japan, Osawa 2-21-1, Mitaka, Tokyo 181-8588, Japan}

\affil{$^{6}$ Institute of Astronomy, Faculty of Science, University of Tokyo, Osawa 2-21-1, Mitaka,
Tokyo, 181-0015, Japan}

\affil{$^{7}$ Institute of Astrophysics and Planetary Sciences, Ibaraki University, Bunkyo 2-1-1, Mito
310-8512, Japan}

\affil{$^{8}$ Institute of Space and Astronoutical Science, Japan Aero space Exploration Agency, Yoshinodai 3-1-1, Sagamihara, Kanagawa 229-8510, Japan}

\affil{$^{9}$ Department of Astronomy, University of Massachusetts, Amherst, MA 01003, USA}

\affil{$^{10}$ Center for Astrophysics and Space Astronomy, University of Colorado, Boulder, CO 80309, USA}

\affil{$^{11}$ Instituto Nacional de Astrof\'{i}sica, \'{O}ptica y Electr\'{o}nica (INAOE), Aptdo. Postal 51 y 216, 72000 Puebla, M\'{e}xico}

\affil{$^{12}$ Department of Physics and Astronomy, University of Pennsylvania, Philadelphia, PA 19104, USA}

\affil{$^{13}$ Department of Earth and Planetary Sciences Tokyo Institute of Technology, 2-12-1, Ookayama, Meguro-ku, Tokyo 152-8551, Japan}

\email{yoshito.shimajiri@nao.ac.jp}



%

\KeyWords{stars:formation ---
ISM: molecules ---
ISM: cloud
} 

\maketitle

\begin{abstract}  
 We present new, wide and deep images in the 1.1 mm continuum and the $^{12}$CO ($J$=1--0) emission toward the northern part of the Orion A Giant Molecular Cloud (Orion-A GMC). 
The 1.1 mm data were taken with the AzTEC camera mounted on the Atacama Submillimeter Telescope Experiment (ASTE) 10 m telescope in Chile, and the $^{12}$CO ($J$=1--0) data  were with the 25 beam receiver (BEARS) on the NRO 45 m telescope in the On-The-Fly (OTF) mode. 
The present AzTEC observations are the widest $(\timeform{1.D7}$ $\times$ $\timeform{2.D3}$, corresponding to 12 pc $\times$ 17 pc) and the highest-sensitivity ($\sim$9 mJy beam$^{-1}$) 
1.1 mm dust-continuum imaging of the Orion-A GMC with an effective spatial resolution of $\sim$ 40$\arcsec$. 
The $^{12}$CO ($J$=1--0) image was taken over the northern $\timeform{1D.2} \times\timeform{1D.2}$ (corresponding 9 pc $\times$ 9 pc) area with a sensitivity of 0.93 K in $T_{\rm MB}$, a velocity resolution of 1.0 km s$^{-1}$, and an effective spatial resolution of 21$\arcsec$. 
With these data, together with the MSX 8 $\mu$m, Spitzer 24 $\mu$m and the 2MASS data, 
we have investigated the detailed structure and kinematics of molecular gas associated with the Orion-A GMC and have found evidence for interactions between molecular clouds and the external forces that may trigger star formation. 
Two types of possible triggers were revealed; 
1) Collision of the diffuse gas on the cloud surface, particularly at the eastern side of the OMC-2/3 region, and
2) Irradiation of UV on the pre-existing filaments and dense molecular cloud cores. 
Our wide-field and high-sensitivity imaging have provided the first comprehensive view of the potential sites of triggered star formation in the Orion-A GMC.

\end{abstract}

\section{Introduction}

Giant Molecular Clouds (GMCs) are main sites of star formation. 
There are approximately three thousand GMCs in our Galaxy, with a size of 10 - 80 pc and a mass of 10$^{4}$ - 5 $\times$ 10$^{6}$ M$_{\odot}$ \citep{Solomon77}. 
While only low-mass stars form in dark clouds \citep{Myers95,Momose98, Saito99, Takakuwa04}, in GMCs both low-mass and high-mass stars form \citep{Megeath94,Homeir05,Yonekura05}. 
There are many uncertain aspects of star formation in GMCs, such as the core mass function (CMF), its link to the initial mass function (IMF),  massive star formation, and cluster formation. 
Hence, studies of star formation in GMCs are imperative to the comprehensive understanding of star formation.   

Theoretical studies of massive star formation and cluster formation have proposed that external effects on molecular clouds, such as ultraviolet (UV) radiation from HII regions, supernova remnants (SNRs), molecular outflows, and cloud-cloud collisions are required to trigger the massive-star \citep{Elmegreen98}  and cluster formation \citep{Whitworth94, Hosokawa05, Nakamura07}. 
Furthermore, it is proposed that the shape of the CMF could also be affected
by the presence of those external effects \citep{Zinnecker07}. 
Hence, it is essential to unveil the global picture of those external effects
on molecular gas in GMCs. 
It is, however, not yet clear how ubiquitous those external effects
and the triggered star formation in GMCs are.
Unbiased and extensive mapping observations of GMCs
in molecular and dust continuum emission,
covering an entire GMC and resolving
individual cores with a typical size of $\sim$ 0.1 pc,
and the comparison with the images at other wavelengths,
are required to approach the global picture of star formation in GMCs.

The Orion-A GMC, with an extent of $\sim$ $\timeform{10D}$ (corresponding to $\sim$ 72 pc) in the southern part of Orion constellation, is the nearest ( $d$= 400 pc; \cite{Menten07, Sandstrom07, Hirota08}) GMC and one of the well studied regions in star formation studies \citep{Tatematsu93, Lis98, Tatematsu98, Johnstone99, Aso00, Stanke02, Tsujimoto02, Takahashi08, Takahashi09, Davis09}. 
Many authors have proposed that the Orion-A GMC is externally affected, both at small ($\sim$ 0.1 pc) and large scales ($\sim$ several $\times$ 10 pc), by molecular outflow, ionizing radiation, SNRs, and stellar winds generated by the Orion OB association, from its morphological structure, velocity structure, and the stellar ages \citep{Cowie79, Bally87, Sandell01,Yokogawa03, Wilson05, Lee09}. 
OB stars embedded in the Orion-A GMC form the HII region, M 42, 43, and NGC 1977 \citep{Goudis82}. 
Molecular outflows driven from OMC-2/FIR 3 and FIR 6c interact with the surrounding dense cores, 
which triggers gravitational instabilities in the dense cores and the next generation star formation \citep{Shimajiri08, Shimajiri09}. 
These facts imply that the Orion-A GMC is one of the most suitable GMCs for an unbiased extensive study of triggered star formation. 

With the advent of multi-beam receivers at millimeter wavelengths, wide-field mapping observations over several square degrees at an angular resolution of several $\times$ 10$\arcsec$ are now feasible. 
In this paper, we present a wide, sensitive 1.1-mm dust-continuum and $^{12}$CO ($J$=1--0) line imaging of the northern part of the Orion A GMC with the AzTEC camera mounted on the ASTE telescope and BEARS mounted on the Nobeyama (NRO) 45-m telescope, respectively.
The aim of this paper is to reveal the overall structure of the Orion-A GMC from a cloud ($\ge$ 10 pc) to a core scale ($\le$ 0.1 pc ) 
with the wide coverage and high sensitivity, 
and specifically, to unveil the global picture of the external triggers of star formation in the GMC 
from the comparison with the near-infrared (NIR) and mid-infrared (MIR) images. 
The detailed physical processes of individual triggered star formation will be discussed in the forthcoming papers. 
The 1.1 mm dust-continuum emission is optically thin and traces the total mass of molecular gas in cores and envelopes \citep{Enoch06}. The $^{12}$CO ($J$=1--0) emission traces low-density gas ($\le$ 10$^3$ cm$^{-3}$) and is usually optically thick and easily thermalized. 
Hence, the $^{12}$CO emission traces global structures of GMCs, and the peak intensities of the $^{12}$CO emission at the emission ridge often reflect kinetic temperature of molecular gas.

This paper is organized as follows: in section 2, the NRO 45-m and AzTEC/ASTE observations are described. In section 3, we present results of the AzTEC 1.1 mm continuum and the NRO $^{12}$CO ($J$=1--0) mapping of the Orion-A GMC. 
In section 4, we focus on the possible external triggers of the star formation in the Orion-A GMC. 
First, we describe the possibility that the entire Orion-A GMC is influenced over $\sim$ several $\times$ 10 pc scale by the external effect. 
Then, we show the region irradiated by the UV radiation from OB stars embedded in the Orion-A GMC over a few pc-scale.  
In section 5, we summarize this paper.

\section{Observations}

\subsection{AzTEC/ASTE Observations}
From October to December 2008, we carried out 1.1 mm dust-continuum observations toward the $\timeform{1.D7}$ $\times$ $\timeform{2D.3}$ region in the northern part of the Orion A molecular cloud with the AzTEC camera \citep{Wilson08} equipped in the ASTE 10 m telescope \citep{Ezawa04, Kohno04} located at Pampa la Bola (altitude = 4800 m), Chile. 
The weather conditions during our observing period were moderate and the typical opacity at 225 GHz was in the range of 0.1 - 0.2.
The AzTEC camera mounted on the ASTE telescope is an 144-element bolometric camera tuned to operate in the 1.1 mm atmospheric window, and
provides an angular resolution of 28$\arcsec$ in full width at half maximum (FWHM)\citep{Wilson08}. 
Observations were performed in the raster scan mode 
toward a pair of 100$\arcmin$ $\times$ 100$\arcmin$ fields, centered on ($\alpha_{J2000}$,$\delta_{J2000}$)=($\timeform{5h35m14s.5}, \timeform{-5D22'30".4}$) and ($\timeform{5h35m 14s.5}, \timeform{-6D00'30".0}$). 
Each field was observed several times with azimuth and elevation scans. 
The separation among scans was adopted to be 117$\arcsec$, which is a quarter of the AzTEC field of view (FoV;$\sim$ $\timeform{7'.8}$). 
A scanning speed of the telescope was 250$\arcsec$ sec$^{-1}$. 
Totally 30 individual maps in the entire field with an integration time of 26 minutes were taken, and those maps were averaged to produce the final map with the total integration time of 26 minutes $\times$ 30 maps = 13 hours. 
The telescope pointing was checked every 2 hours by observing a quasar 0539-057 over a 4 $\times$ 4 arcmin$^{2}$ region. 
The derived pointing offsets were linearly interpolated along the time sequence and the interpolated pointing offset was applied to each target map. 
The pointing uncertainty of the AzTEC map is estimated to be better
than 2$\arcsec$. We observed Uranus as a flux calibrator twice per night.
By observing Uranus with each AzTEC detector element, we measured the flux conversion factor (FCF) from the optical loading value (in watts) to the source flux (in Jy beam$^{-1}$) 
for each detector element. The principal component analysis (PCA, \cite{Scott08}) cleaning method was applied to remove the atmospheric noise. Details of the flux calibration are described by \citet{Wilson08} and \citet{Scott08}.
Since the PCA method does not have sensitivity to extended sources, 
we apply the iterative mapping method (FRUIT, \cite{Liu10}) to recover
the extended components. 
The noise level is $\sim$ 9 mJy beam$^{-1}$ in the central region and $\sim$ 12 mJy beam$^{-1}$ in the outer edge, 
the effective beam size is $\sim$ 40$\arcsec$, after the FRUIT imaging.

We describe the performance of FRUIT in appendix 1.

\subsection{NRO 45m $^{12}$CO ($J$=1--0) Line Observations}
From December 2007 to May 2008, we carried out $^{12}$CO ($J$=1--0; 115.271204 GHz) line  observations toward the $\timeform{1.D2}$ $\times$ $\timeform{1D.2}$ region in the northern part of the Orion A molecular cloud with the 25-element focal plane receiver BEARS equipped in the 45 m telescope at NRO. 
At 115 GHz, the telescope has a FWHM beam size of 15$\arcsec$ 
and a main beam efficiency of 32 \%.
The beam separation of the BEARS is $\timeform{41".1}$ on the plane of the sky \citep{Sunada00, Yamaguchi00}. 
At the back end, we used 25 sets of 1024
channel auto-correlators (ACs) which have the bandwidth of 32 MHz 
and the frequency resolution of 37.8 kHz \citep{Sorai00}.
The frequency resolution
corresponds to the velocity resolution of $\sim$ 0.1 km s$^{-1}$ at 115 GHz.
During the observations, the system
noise temperatures were in the range between 250 and 500 K
in a double sideband. The standard chopper
wheel method \citep{Kutner81} 
was used to convert the output signal into the antenna temperature $T_{\rm A}^{*}$ (K), corrected for the atmospheric
attenuation. The telescope pointing was checked every 1.5 hours
by observing a SiO maser source, Ori-KL, and was
better than 3$\arcsec$ during the whole observing period. 
To correct the gain variation among the 25 beams of the BEARS, we used
calibration data obtained from the observations toward Orion IRC 2 using another SIS receiver, S100, with a single sideband (SSB)
filter and acousto-optical spectrometers (AOSs) as a back end.

Our mapping observations were made with the OTF mapping
technique \citep{Sawada08}. We used the emission-free area near
($\sim$ $\timeform{2D}$ away) the observed area as the off position. 
We obtained an OTF map with a scanning direction both along the RA and DEC 
over the $\timeform{1.D2}$ $\times$ $\timeform{1D.2}$ region 
and combined them into a single map, in order to reduce the
scanning effects. 
We adopted a convolutional scheme with a spheroidal function
\citep{Sawada08} to calculate the intensity at each grid point 
of the final map-cube data with a spatial grid
size of $\timeform{7".5}$, half of the beam size. The resultant effective resolution
is 21$\arcsec$. The rms noise level of the final map is
0.94 K in $T_{\rm MB}$ (1 $\sigma$) at a velocity resolution of 1.0 km s$^{-1}$.

We summarize the parameters for the 1.1 mm continuum and the $^{12}$CO ($J$=1--0) line observations in Table \ref{obs}.

\section{Results}
\subsection{1.1 mm Dust Continuum Emission}

\begin{figure}
   \begin{center}
      \FigureFile(150mm,150mm){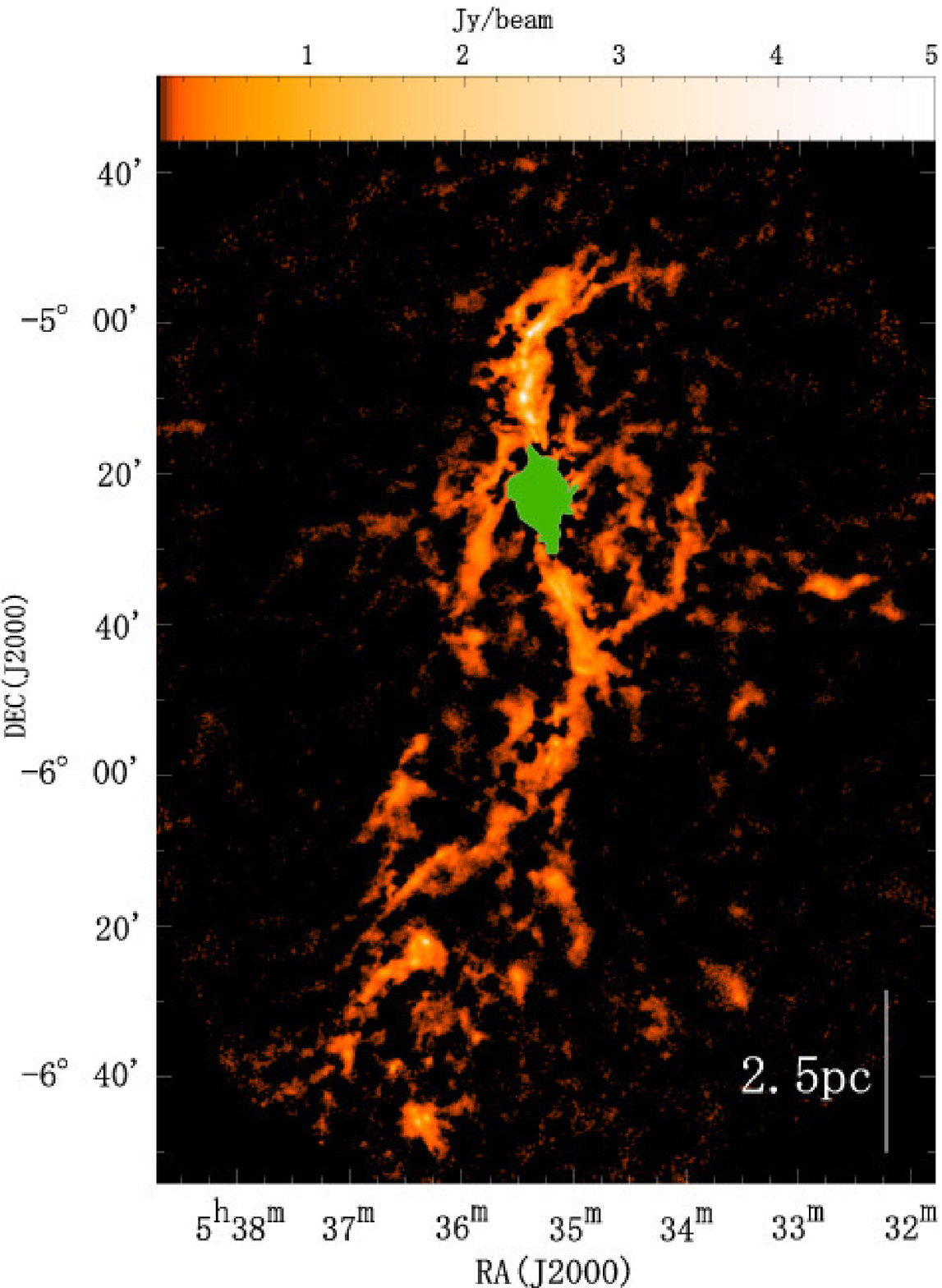}      
   \end{center}
 \footnotesize
   \caption{AzTEC/ASTE 1.1 mm dust continuum map. The rms noise level is 9 mJy beam$^{-1}$ in the central  region and 12 mJy beam$^{-1}$ on the outer edge. The central Orion-KL region has been masked out for this analysis, because the continuum emission around Orion-KL is too bright to be reconstructed as an accurate structure with the AzTEC data-reduction technique.  
}\label{aztec}
\end{figure}

Figure \ref{aztec} shows the distribution of the 1.1 mm dust-continuum emission
in the northern part of the Orion A molecular cloud.
Previous observations of the northern part of the Orion A molecular cloud
in dust-continuum emissions at 850 $\mu$m \citep{Johnstone99, Nutter07}, 1.2 mm \citep{Davis09}, and 1.3 mm \citep{Chini97} 
have revealed an ``integral-shape'' filamentary feature
from the south to the north of Ori-KL ($\alpha_{J2000}$, $\delta_{J2000}$ = $\timeform{5h35m14s.5},
\timeform{-5D22'30".4}$), where the northern end, region surrounding Ori-KL, and the southern region of Ori-KL in the integral-shape filament are called OMC-2/3, OMC-1, and OMC-4, respectively.
The global feature seen in our 1.1 mm dust-continuum map is consistent with that found
in the previous observations, that is, the integral-shape filament
along the north-south direction (named as "Integral-shape filament" in  Figure \ref{aztec2}).
The central Orion-KL region has been masked out, 
because around the Ori-KL region the continuum emission is too bright to be reconstructed as an accurate structure with the AzTEC observing and data-reduction technique
(see Appendix 1).

Our mapping area extends $\sim$ $\timeform{1D.7}$ along the
eastern and the western directions centered on the integral-shape filament,
and is wider than that of the previous observations.
As a result, we have found new dust-continuum features
in the eastern and the western areas of the integral-shape filament, as described below.

\subsubsection{New Features around the OMC-2/3 Region}

\begin{figure}
   \begin{center}
      \FigureFile(150mm,150mm){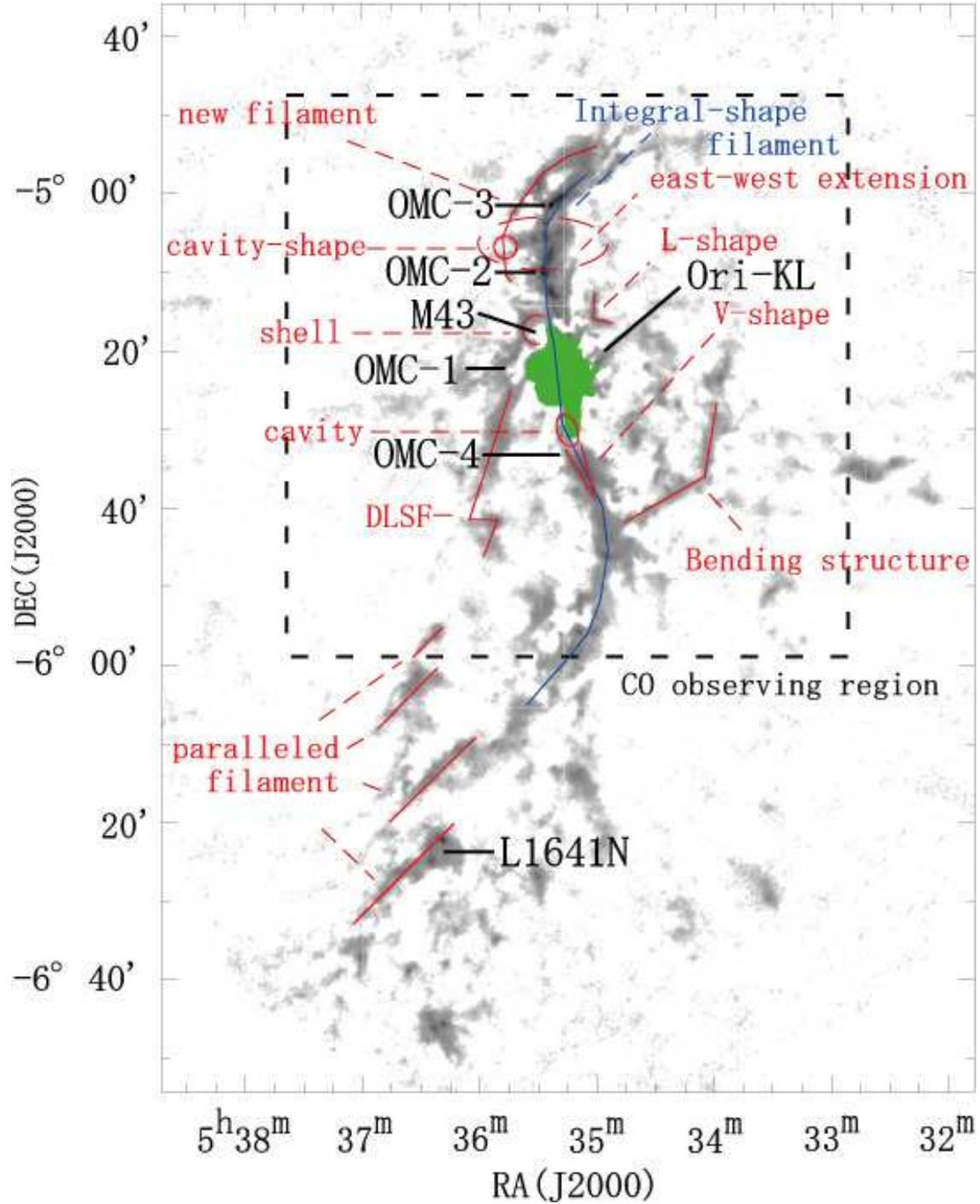}      
   \end{center}
 \footnotesize
   \caption{Map of main 1.1 mm dust continuum features. The details of the features are described in section 3.1. A black box, $\sim$ 1.2 $\times$ $\sim$1.2 degree$^2$ in size, shows the $^{12}$CO ($J$=1--0) observing region. The central Orion-KL region has been masked out for this analysis, because the continuum emission around Orion-KL is too bright to be reconstructed as an accurate structure with the AzTEC data-reduction technique.  
}\label{aztec2}
\end{figure}

Toward the east of the OMC-2/3 region, our high-sensitivity ($\sim$ 9 mJy) 1.1 mm continuum observations
have revealed the presence of another filamentary structure (named as "new filament" in Figure \ref{aztec2}).
In addition, a continuum feature extending almost perpendicular to the integral-shape filament centered on MMS 9 ($\alpha_{J2000}$, $\delta_{J2000}$ = $\timeform{5h35m26s.00}, \timeform{-5D05'42".4}$) was detected clearly at a signal to noise ratio of $\ge$ 50 $\sigma$, although this
feature was not clearly seen in the previous studies
(named as "east-west extension" in Figure \ref{aztec2}
).
In the most eastern side of this structure, a cavity-like structure with a size of 0.4 pc was found at a central position of $\alpha_{J2000}$, $\delta_{J2000}$ = $\timeform{5h35m44s}, \timeform{-5D06'48"}$ (named as "cavity-shape" in Figure \ref{aztec2}).

\subsubsection{Features around the HII regions, M 42 and M 43}

The distribution of the 1.1 mm dust-continuum emission around the HII region, M 43,
which is excited by a B0.5 star, NU Ori ($\alpha_{J2000}$, $\delta_{J2000}$ = $\timeform{5h35m31s.36}, \timeform{-5D16'02".5}$) \citep{Thum78},
shows a shell-like structure with a size of $\sim$ 0.5 pc
(named as ``shell'' in Figure \ref{aztec2}). 
 Toward the south of M 42 (OMC-4 region) the 1.1 mm dust-continuum emission shows a ``V-shape'' structure with a size of $\sim$ 2 pc, which is also seen  in the 850 $\mu$m continuum emission \citep{Johnstone99}, and
a cavity-like structure with a central position of ($\alpha_{J2000}$, $\delta_{J2000}$ = $\timeform{5h35m13s}, \timeform{-5D31'42"}$)(named as ``cavity'' in Figure \ref{aztec2}). Toward the northwest of M 42, the 1.1 mm dust-continuum emission forms an ``L-shape'' structure with a size of $\sim$ 0.5 pc (named as "L-shape" in Figure \ref{aztec2}).

\subsubsection{Filaments in peculiar shapes outside the integral filament}

As well as the integral-shape filament there appears
a branch of the 1.1 mm dust-continuum emission extending toward the southeast of Ori-KL.
The branch forms a filamentary structure with a length of $\sim$ 3 pc along $\alpha_{J2000}$ $\sim$ $\timeform{5h36m}$.
This filament is known as one of photo-dominated regions (PDRs) from the CN line observations \citep{Rod01}
and is named as ``dark lane south filament'' (DLSF, named as ``DLSF'' in Figure \ref{aztec2}).
Furthermore, toward the opposite side of DLSF with respect to the integral-shape filament 
there is another filamentary structure 
with a size of $\sim$ 3 pc, which is bending at the middle
(named as ``Bending structure'' in Figure \ref{aztec2}).

In the southmost region where an active cluster-forming region of L 1641 N \citep{Chen95, Chen96, Stanke07} resides, 
four filamentary structures with a size of $\sim$ 0.5 - 2.0 pc are seen aligned almost in parallel (named as "paralleled filament" in Figure \ref{aztec2}).

\subsection{$^{12}$CO ($J$=1--0) Emission}

\begin{figure}
   \begin{center}
         \FigureFile(160mm,160mm){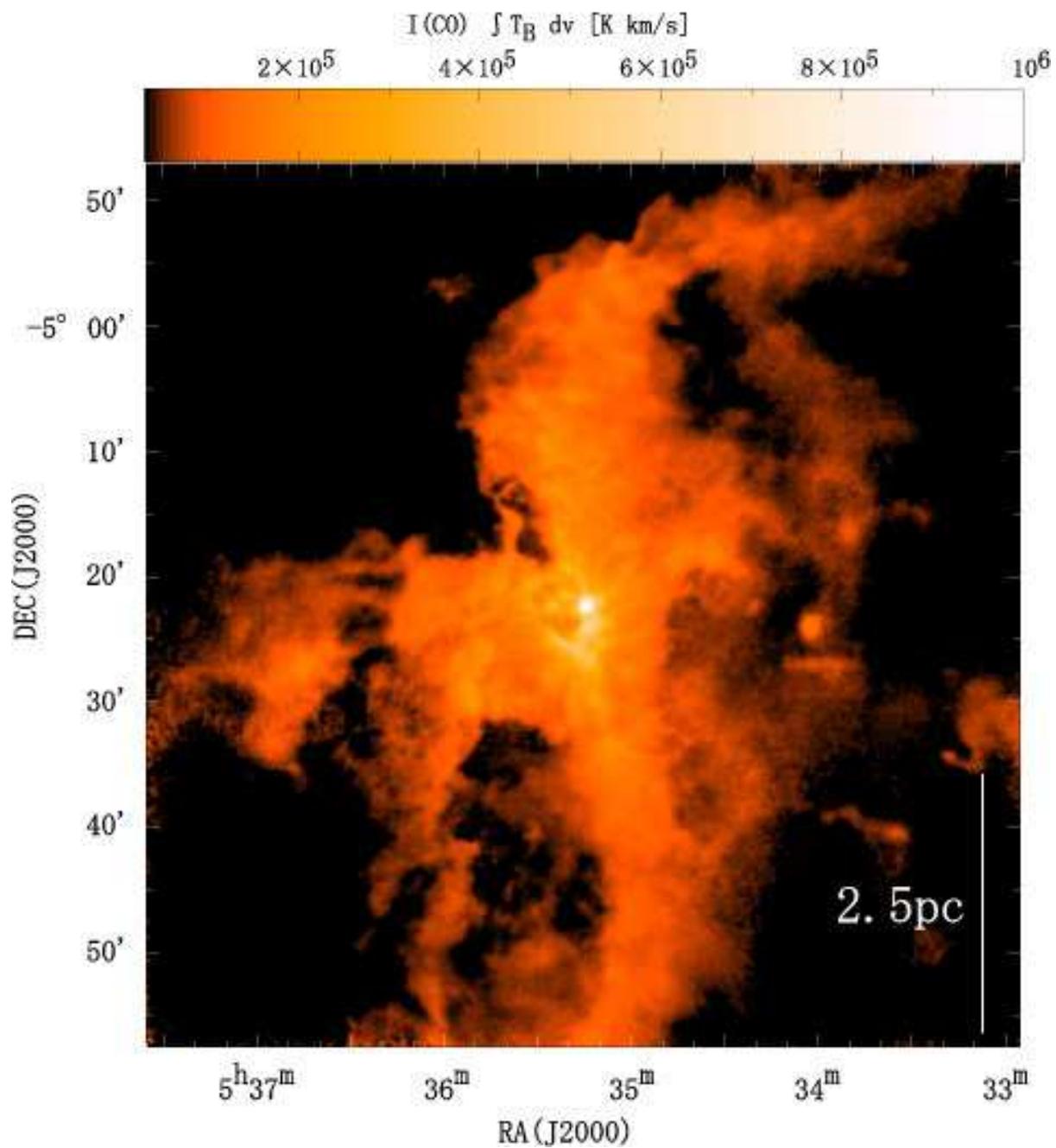}           
   \end{center}
 \footnotesize
   \caption{$^{12}$CO ($J$=1--0) total integrated intensity map at a velocity range of 0.0 - 20.0 km s$^{-1}$. The noise level (1 $\sigma$) is 70.0 K km s$^{-1}$, and the peak intensity is $\sim$ 1.62 $\times$ 10$^{7}$ K km s$^{-1}$.  
}\label{co_alla}
\end{figure}

\begin{figure}
   \begin{center}
          \FigureFile(160mm,160mm,angle=40){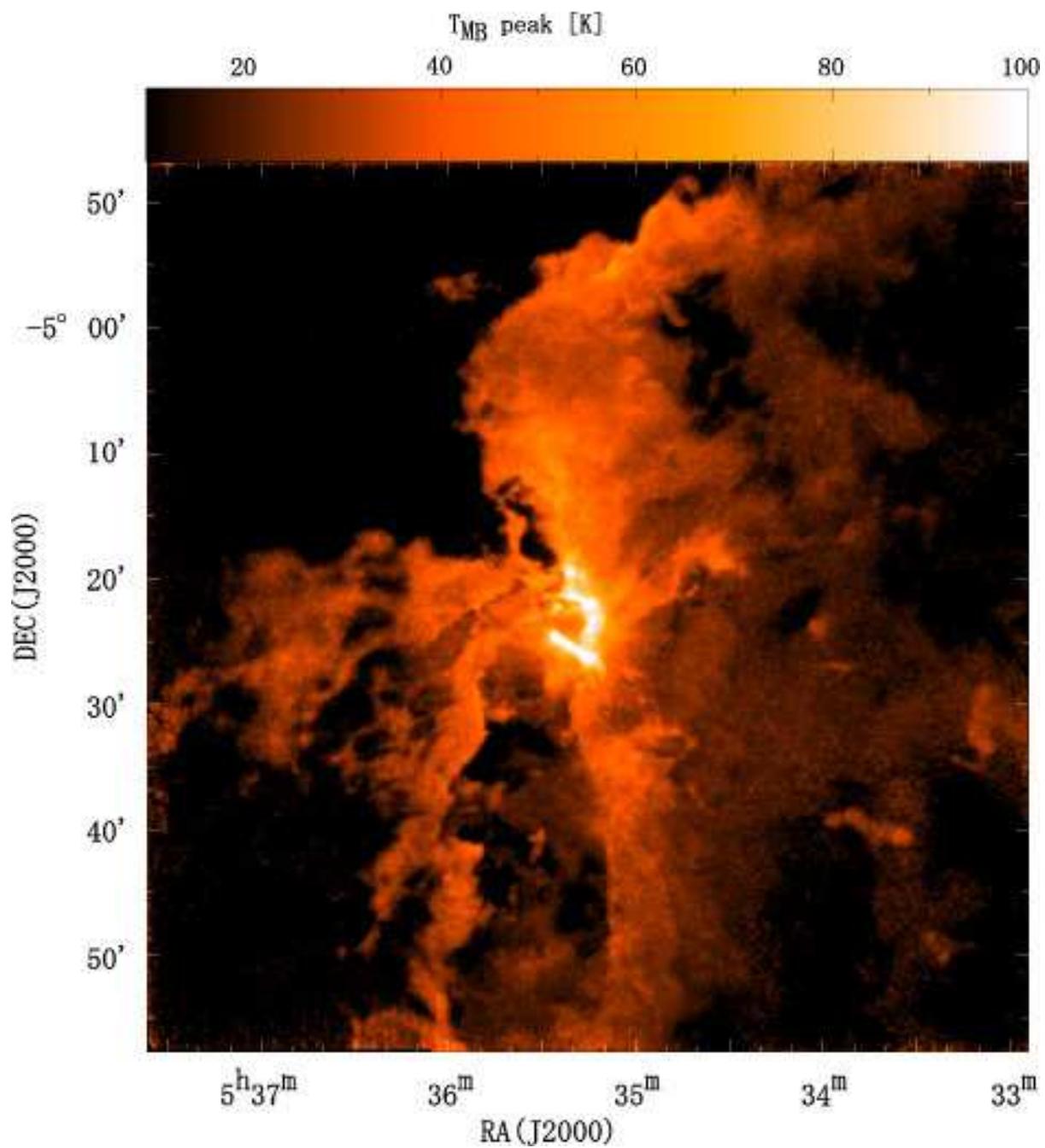} 
   \end{center}
 \footnotesize
   \caption{Peak intensity map in the $^{12}$CO ($J$=1--0) line. The noise level (1 $\sigma$) is 1.4 K at $T_{\rm MB}$ and the peak intensity is $\sim$ 111.7 K at $T_{\rm MB}$.
}\label{co_allb}
\end{figure}

\begin{figure}
   \begin{center}
          \FigureFile(160mm,160mm){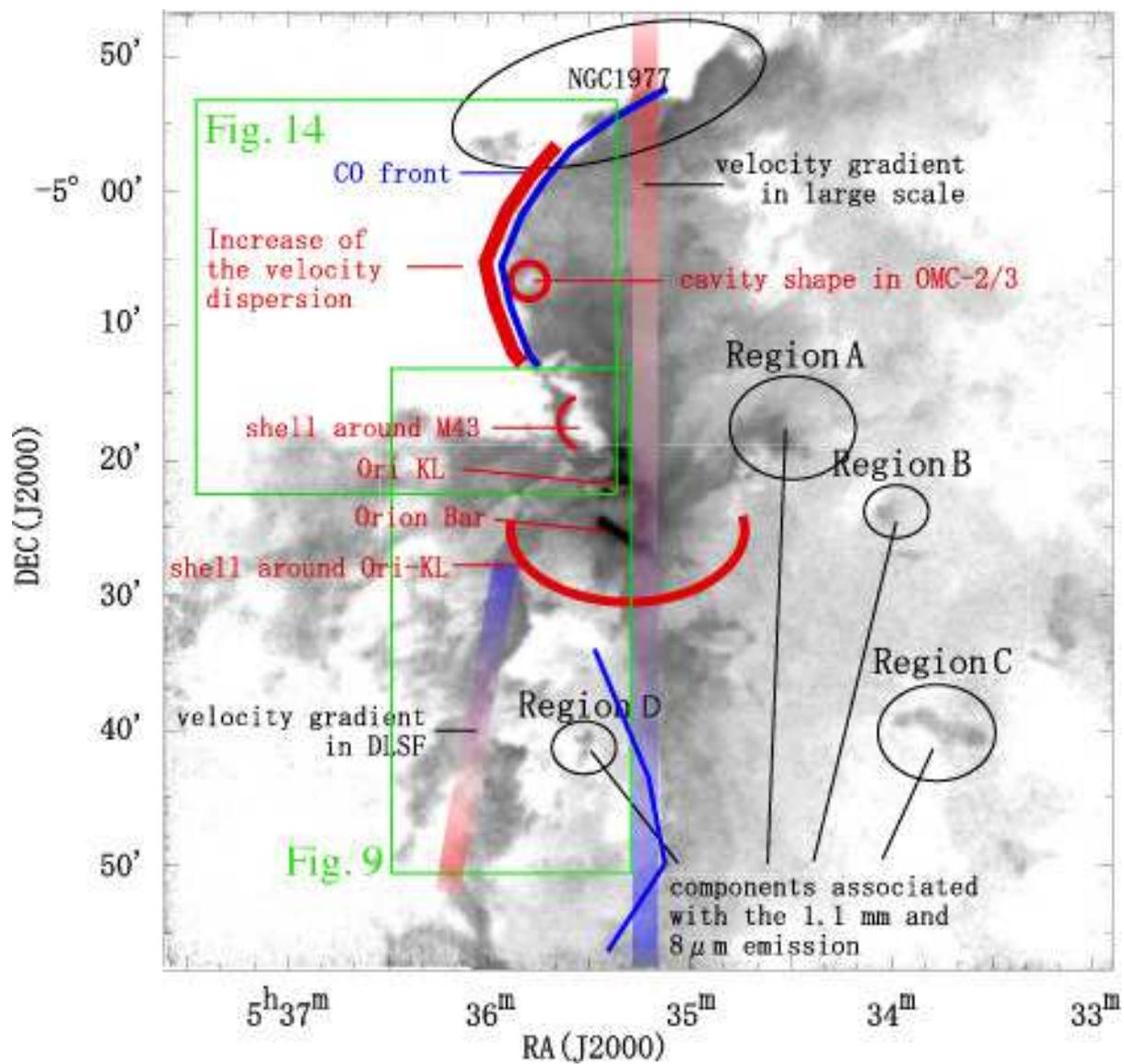}          
   \end{center}
 \footnotesize
   \caption{Map of $^{12}$CO ($J$=1--0) emission features. Grey scale is a peak intensity map in the $^{12}$CO ($J$=1--0) line. The detail of the features are described in section 3.2.     
}\label{co_allc}
\end{figure}

Figure \ref{co_alla}  and \ref{co_allb} show the total integrated and the peak intensity maps
of the $^{12}$CO ($J$=1--0) emission in the northern part of the Orion A molecular cloud,
respectively. The main feature seen in these maps is the same as that in the 1.1 mm continuum map, namely,
integral-shape filament. At the eastern edge of the integral-shape filament there appears
a sharp emission contrast while at the western edge the emission intensity decreases gradually.
Hereafter we call the sharp emission contrast at the eastern edge of the integral-shape filament ``CO front''
(see Figure \ref{co_allc}). 
In the total integrated intensity map the brightest point
at the center corresponds to the location of Ori-KL. Toward the south-east of Ori-KL
there appears a bar-like feature with a length of $\sim$ 0.6 pc,
which is more evident in the peak intensity map. This feature corresponds to ``Orion-Bar''
(see Figure \ref{co_allc}). As well as the $^{12}$CO components associated with the integral-shape filament,
there appears a branch of the $^{12}$CO emission extending toward the southeast from Ori-KL.
This $^{12}$CO structure corresponds to DLSF seen in the AzTEC 1.1 mm dust-continuum
emission (see Figure \ref{co_allc} as well as Figure \ref{aztec2}).
Similarly to CO front, there is a sharp emission contrast
at the western side of DLSF.
In addition to these filamentary features,
a $^{12}$CO emission extending to the east from Orion-KL and a diffuse CO emission at the west
of the integral-shape filament are seen.
These $^{12}$CO emission features are consistent with the features found by the
previous $^{12}$CO ($J$=1--0) observations at
an angular resolution of 45$\arcsec$ and a grid spacing of 45$\arcsec$ \citep{Heyer92}. 
Toward the emission ridge, the peak intensity map of the $^{12}$CO ($J$=1--0) emission
probably shows the temperature structure, under the condition that
the $^{12}$CO ($J$=1--0) emission is optically thick and thermalized.
The temperature around Ori-KL is the highest ($\sim$ 110 K), and the temperature around
NGC 1977 is relatively high ($\sim$ 64 K).

\begin{figure}
   \begin{center}
      \FigureFile(160mm,160mm){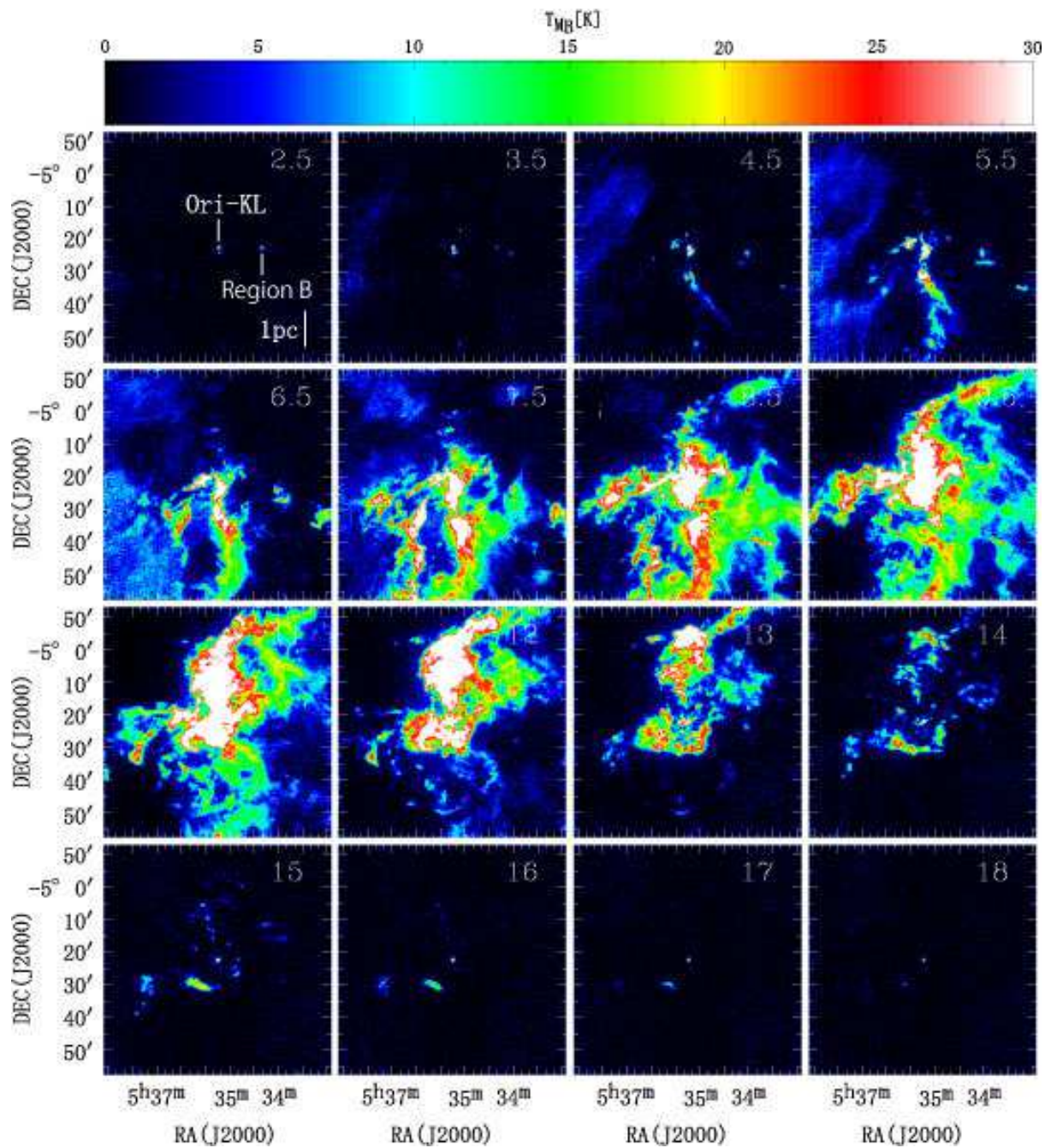} 
   \end{center}
 \footnotesize
   \caption{$^{12}$CO ($J$=1--0) velocity channel maps.
The image size is $\sim$ 1.2 $\times$ 1.2 degree$^2$, and the effective special resolution is $\sim$21$\arcsec$. 
The velocity width of each map is 1.0 km s$^{-1}$, and typical noise level (1 $\sigma$) is 0.93 K in $T_{\rm MB}$. 
}\label{co_channel}
\end{figure}

Figure \ref{co_channel} shows the velocity channel maps of the $^{12}$CO ($J$=1--0) emission
with a velocity width of 1.0 km s$^{-1}$. In the high-velocity range ($V_{\rm LSR}$ = 2.5 - 3.5
$\&$ 15 - 18 km s$^{-1}$), a point-like $^{12}$CO emission associated with Ori-KL is seen
at ($\alpha_{J2000}$, $\delta_{J2000}$) = ($\timeform{5h35m14s.5}, \timeform{-5D22'30".4}$).
From $V_{\rm LSR}$ = 2.5 to 6.5 km s$^{-1}$, there is another compact $^{12}$CO component seen
toward the west of Ori-KL at ($\alpha_{J2000}$, $\delta_{J2000}$) = ($\timeform{5h34m20s.4},
\timeform{-5D22'24".2}$). 
This $^{12}$CO component is associated with the AzTEC 1.1 mm dusty condensation
(named as "Region B", see \S 3.3).
From $V_{\rm LSR}$ = 4.5 km s$^{-1}$ to 14 km s$^{-1}$ the main integral-shape filament
shows a velocity gradient from the south to the north of Ori-KL, as already reported by
previous observations \citep{Bally87, Heyer92, Ikeda07, Tatematsu08}.
On the other hand, 
from $V_{\rm LSR}$ = 5.5 to 8.5 km s$^{-1}$ the $^{12}$CO emission originated from DLSF
shows a velocity gradient from the north to the south (see Figure \ref{co_allc}),
which suggests that the velocity gradient of the $^{12}$CO emission associated with DLSF is opposite
compared to the large scale velocity gradient of the Orion A molecular cloud. 
The central velocity of DLSF ($\sim$ 7 km s$^{-1}$) is also different from that of the integral-shape filament. 
Hence, DLSF is likely to be kinematically distinct from the main cloud component. 
In the velocity range of 12 to 16 km s$^{-1}$, the $^{12}$CO emission forms a shell-like structure
with a size of $\sim$ 2 pc toward the south of Ori-KL (named as "shell around Ori-KL"
in Figure \ref{co_allc}).
\citet{Bally87} and \citet{Heyer92} also found this shell-like
structure around Ori-KL in the $^{13}$CO ($J$=1--0) and the $^{12}$CO ($J$=1--0) lines observations.

As well as these emission-ridge components,
in the velocity range of 3.5 - 7.5 km s$^{-1}$ there is extended diffuse CO component
at the eastern region of the integral-shape filament.
The peak intensity and the velocity width of the diffuse CO component 
are $\sim$ 5 K and $\sim$ 5 km s$^{-1}$, respectively. The origin of this rather featureless and
extended $^{12}$CO emission is likely to be different from that of the ridge components. 
\citet{Sakamoto97} also found the diffuse CO component with a velocity width of $\sim$ 7 km s$^{-1}$ and a peak intensity of $\sim$ 5 K in the east of the L 1641 N region, which locates in the southern part of the AzTEC 1.1 mm continuum map (Figure \ref{aztec}) and out of our $^{12}$CO observing region. 
The diffuse CO component in the east of the L 1641 N region found by \citet{Sakamoto97} is  probably the same component as that in the east of the OMC-2/3 region and in the east of DLSF. 
 Their and our results suggest that the diffuse CO component distributes along the east of the integral-shape filament over the extent of $\ge$ 14 pc.
We will discuss the origin of the diffuse CO component in more detail.

Figure \ref{co_alla}  shows the mean velocity map of the $^{12}$CO ($J$=1--0) emission.
The large-scale velocity gradient from the south to north can be confirmed in this mean 
velocity map as already shown in the channel maps (see Figure \ref{co_channel}). 
Figure \ref{co_allb} shows the velocity dispersion map of the $^{12}$CO ($J$=1--0) emission. 
The velocity dispersion map shows the increase of the 
velocity dispersion toward the east of CO front in the OMC-2/3 region (see Figure \ref{co_allc}).
Moreover, there is a shell-like structure around Ori-KL in the velocity dispersion map
(see Figure \ref{co_allc}).

\subsection{Comparison Among the AzTEC 1.1 mm, $^{12}$CO peak Intensity, 2MASS, and the MSX 8 $\mu$m Map}

\begin{figure}
   \begin{center}
         \FigureFile(140mm,100mm){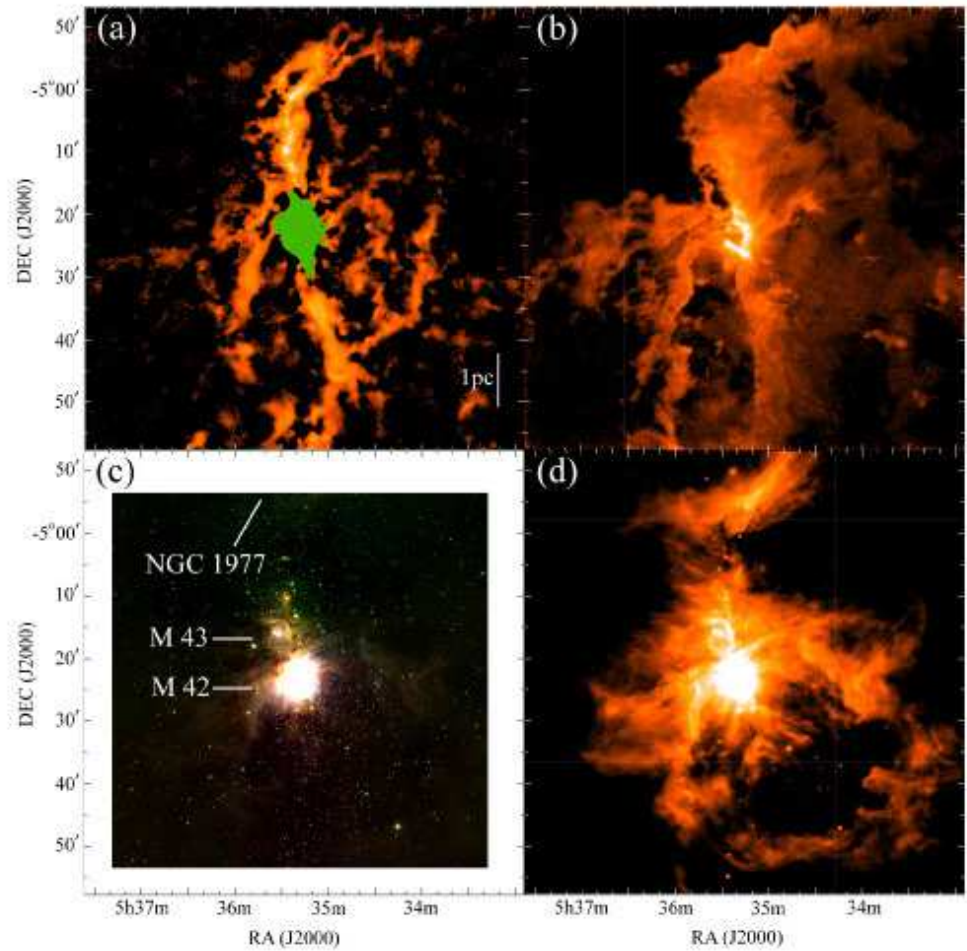} 
   \end{center}
 \footnotesize
   \caption{(a) AzTEC/ASTE 1.1 mm dust-continuum image. The central Orion-KL region has been masked out for this analysis, because the continuum emission around Orion-KL is too bright to be reconstructed as an accurate structure with the AzTEC data-reduction technique.   
(b) peak intensity map in the $^{12}$CO ($J$=1--0) line (c) 2MASS image obtained from the archive of IPAC image mosaic service. (d) MSX 8 $\mu$m image obtained from the archive of IRAC  Services. }\label{multi}
\end{figure}

Figure \ref{multi} (a)-(d) show the AzTEC 1.1 mm, $^{12}$CO ($J$=1--0) peak intensity, 2MASS, and the midcourse space experiment (MSX) 8 $\mu$m maps in the $^{12}$CO observing region, respectively. 
The 8 $\mu$m emission mostly arises from polycyclic aromatic hydrocarbons (PAHs) and has been observed
at the boundary of the HII regions \citep{Zavagno06}. Recent laboratory experiments have shown that PAHs are produced on the dust grains irradiated by the FUV radiation \citep{Bernstein99}. Hence, presence of the strong 8 $\mu$m emission suggests presence of the FUV radiation from HII regions. 
While the 1.1 mm and the $^{12}$CO images show the distribution of the molecular gas, 
the 2MASS and the 8$\mu$m images show three bright nebulae, M 42, M 43, and NGC 1977.  Although the millimeter and infrared emission distributions appear quite distinct from each other, 
detailed comparisons among the 1.1 mm, $^{12}$CO, and the 8 $\mu$m images show their spatial correlations in several regions. 

\begin{figure}
   \begin{center}
         \FigureFile(140mm,100mm){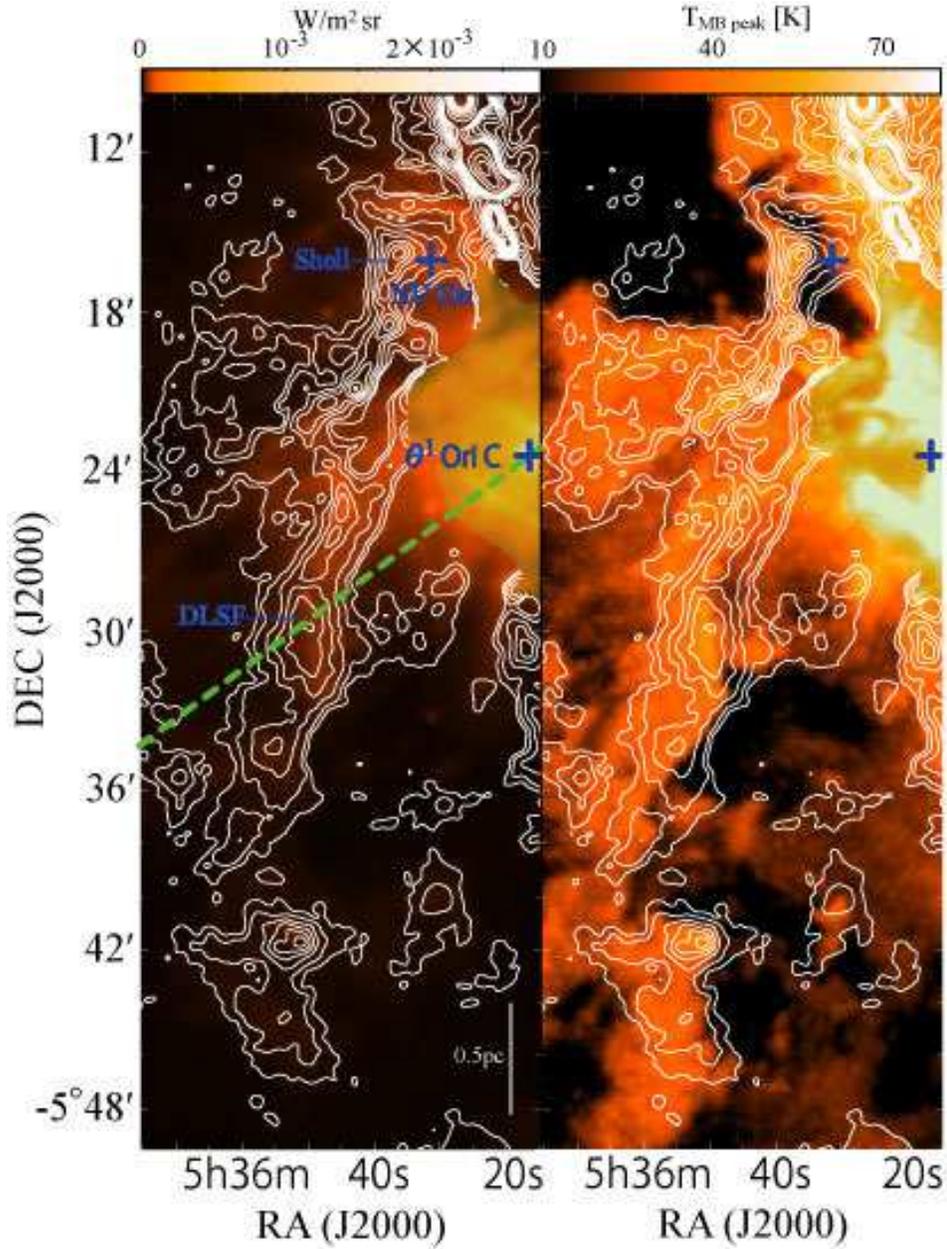}          
   \end{center}
 \footnotesize
   \caption{(a) AzTEC 1.1 mm contours overlaid on a MSX 8 $\mu$m map in the M 43 and DLSF region. (b) AzTEC 1.1 mm contours overlaid on a peak intensity map in the $^{12}$CO ($J$=1--0) line in the M 43 and DLSF region. 
Contour levels start from 5 $\sigma$ with an interval of 10 $\sigma$ for a range of 10 - 100 $\sigma$ and 50 $\sigma$ for a range from 100 $\sigma$. The crosses show the position of NU Ori and $\theta^{1}$ Ori C, respectively. 
The green dash line shows a cut line of the intensity profiles shown in Figure \ref{int_profile}.  
}\label{pdr}
\end{figure}

Figure \ref{pdr} (a) and (b) compare the 1.1 mm continuum distribution to the MSX 8 $\mu$m and the $^{12}$CO ($J$=1--0) emission distributions in the M 43 and DLSF region. 
In this region there are two HII regions, M 42 and M 43, excited by $\theta$$^1$ Ori C and NU Ori, respectively. 
There are systematic trends of the 8 $\mu$m, 1.1 mm, and, the $^{12}$CO emission distributions as a function of the distance from the HII regions. Toward M 43, there is a shell-like structure in the $^{12}$CO, 1.1 mm, and the 8 $\mu$m emissions, however, the radius of the shell decreases systematically from the $^{12}$CO, 1.1 mm, and the 8 $\mu$m emissions. 
DLSF is likely to be irradiated by $\theta$$^1$ Ori C and  the $^{12}$CO, 1.1 mm, and  the 8 $\mu$m emissions show systematic emission distributions as a function of the distance from $\theta$$^1$ Ori C; the $^{12}$CO emission distribution is located farthest and the 8 $\mu$m closest. 

We have also identified other four regions with the systematic $^{12}$CO, 1.1 mm, and the 8 $\mu$m emissions distribution. These regions are labeled as Region A-D, and the locations of Region A-D are shown in Figure 
\ref{co_allc}. 

\begin{figure}
   \begin{center}
         \FigureFile(140mm,100mm){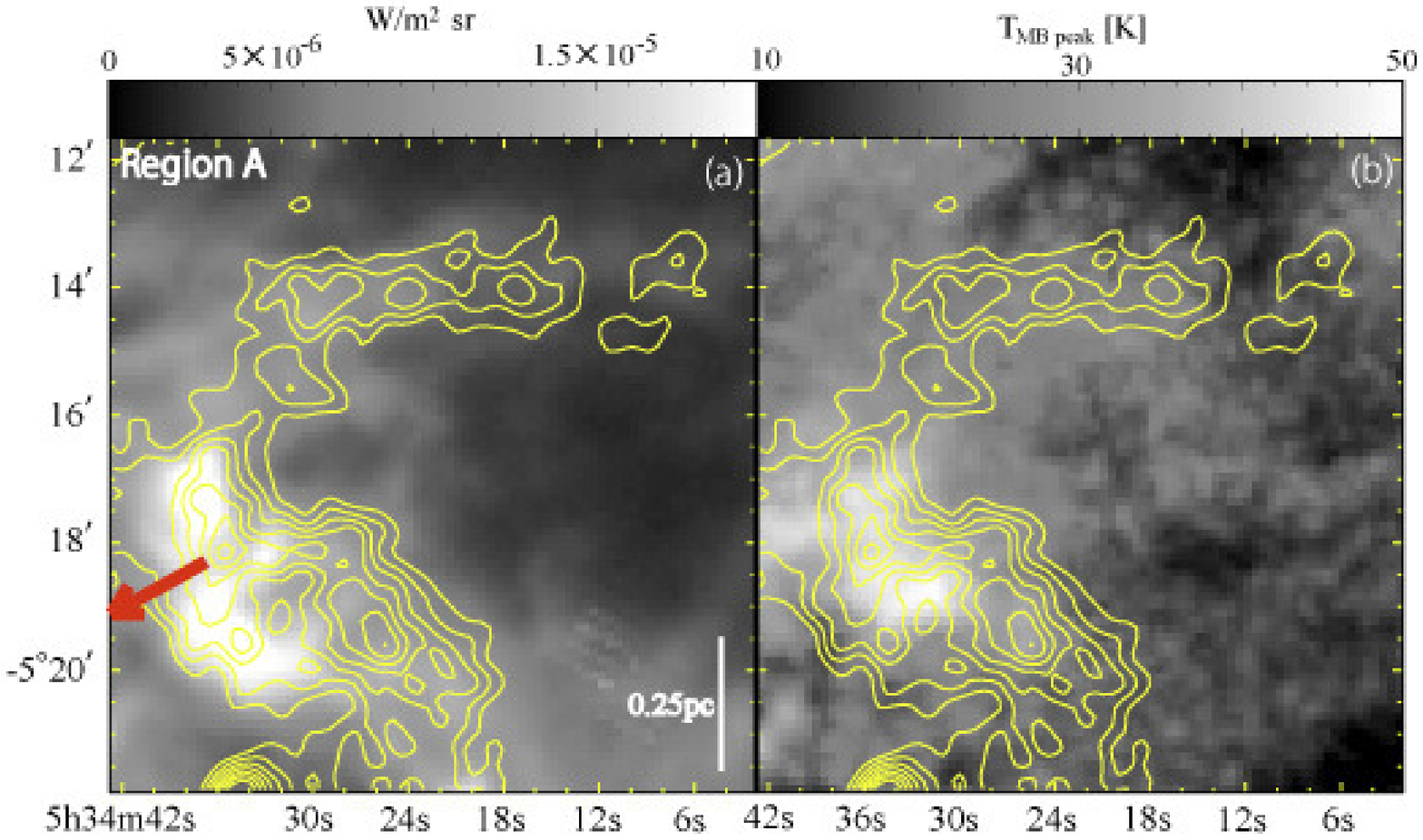}
   \end{center}
 \footnotesize
   \caption{AzTEC/ASTE 1.1 mm contours overlied on a MSX 8 $\mu$m map and peak intensity map in the $^{12}$CO ($J$=1--0) line in Region A. 
Contour levels start from 10 $\sigma$ with an interval of 10 $\sigma$ for a range of 10 - 150 $\sigma$ and 50 $\sigma$ for a range from 150 $\sigma$. 
}\label{aztec_8A}
\end{figure}

\begin{figure}
   \begin{center}
         \FigureFile(140mm,100mm){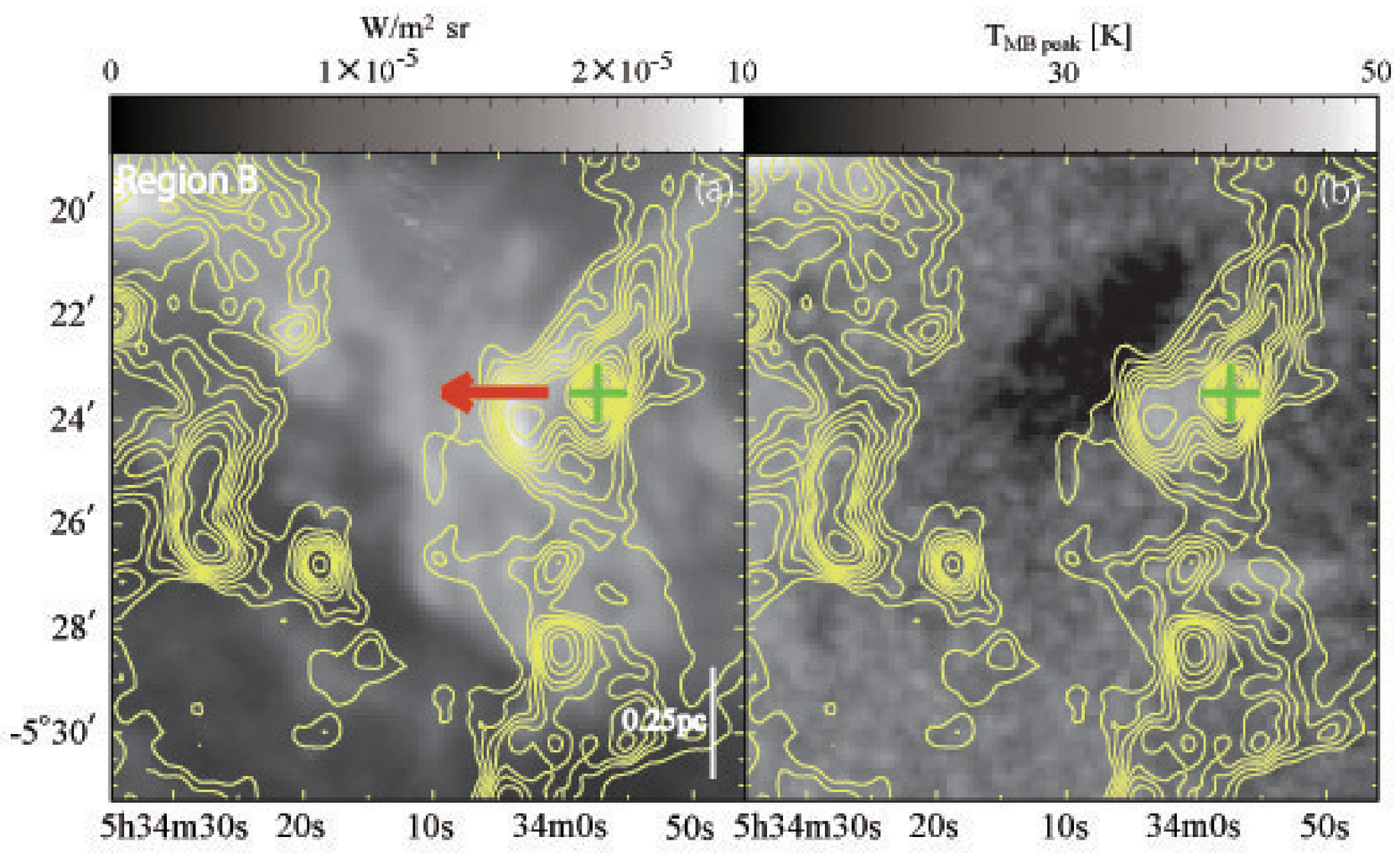}
   \end{center}
 \footnotesize
   \caption{AzTEC/ASTE 1.1 mm contours overlied on a MSX 8 $\mu$m map and peak intensity map in the $^{12}$CO ($J$=1--0) line in Region B. 
Contour levels start from 10 $\sigma$ with an interval of 10 $\sigma$ for a range of 10 - 150 $\sigma$ and 50 $\sigma$ for a range from 150 $\sigma$.
}\label{aztec_8B}
\end{figure}

\begin{figure}
   \begin{center}
         \FigureFile(140mm,100mm){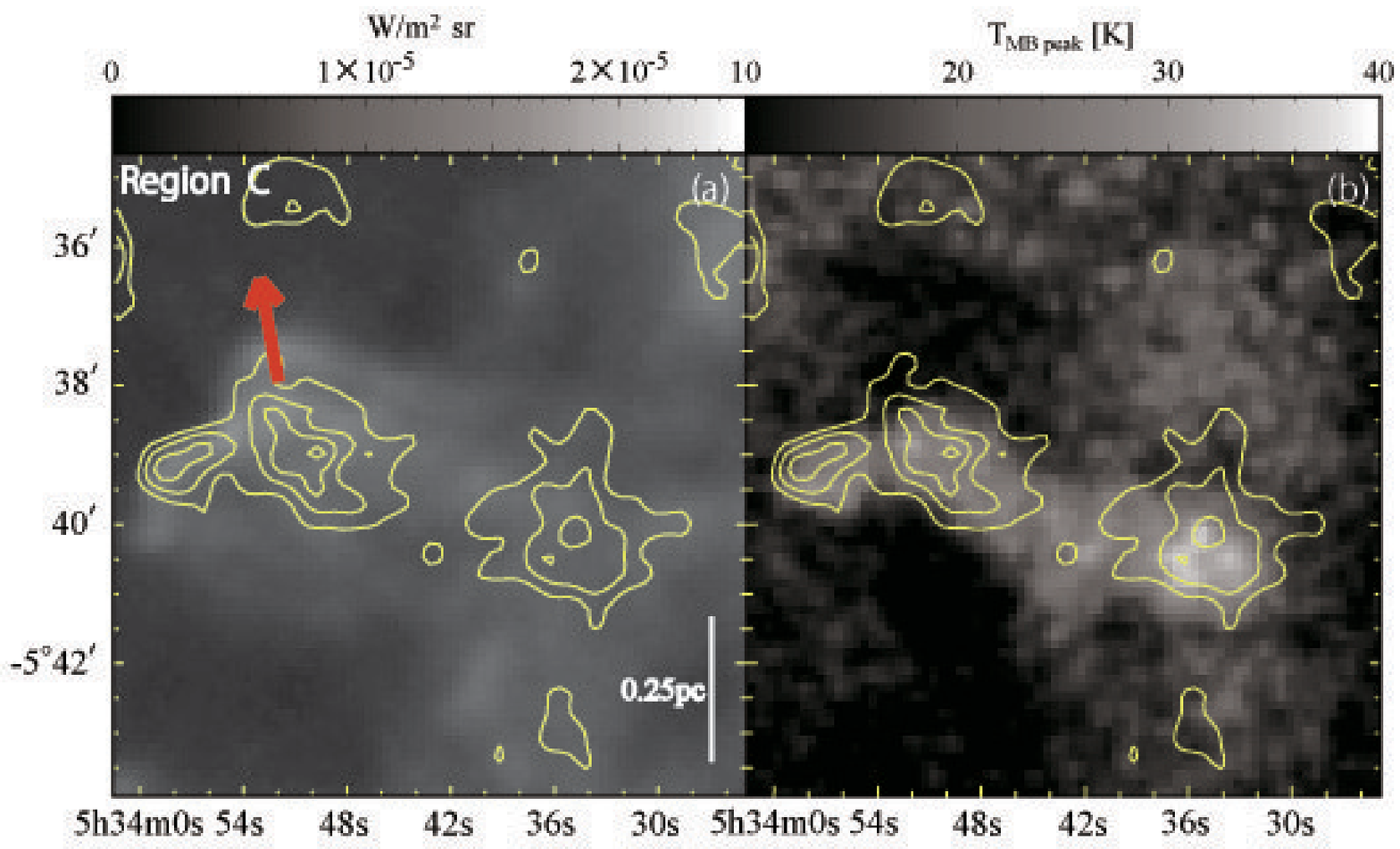}
   \end{center}
 \footnotesize
   \caption{AzTEC/ASTE 1.1 mm contours overlied on a MSX 8 $\mu$m map and peak intensity map in the $^{12}$CO ($J$=1--0) line in Region C. 
Contour levels start from 10 $\sigma$ with an interval of 10 $\sigma$ for a range of 10 - 150 $\sigma$ and 50 $\sigma$ for a range from 150 $\sigma$.
}\label{aztec_8C}
\end{figure}

\begin{figure}
   \begin{center}
         \FigureFile(140mm,100mm){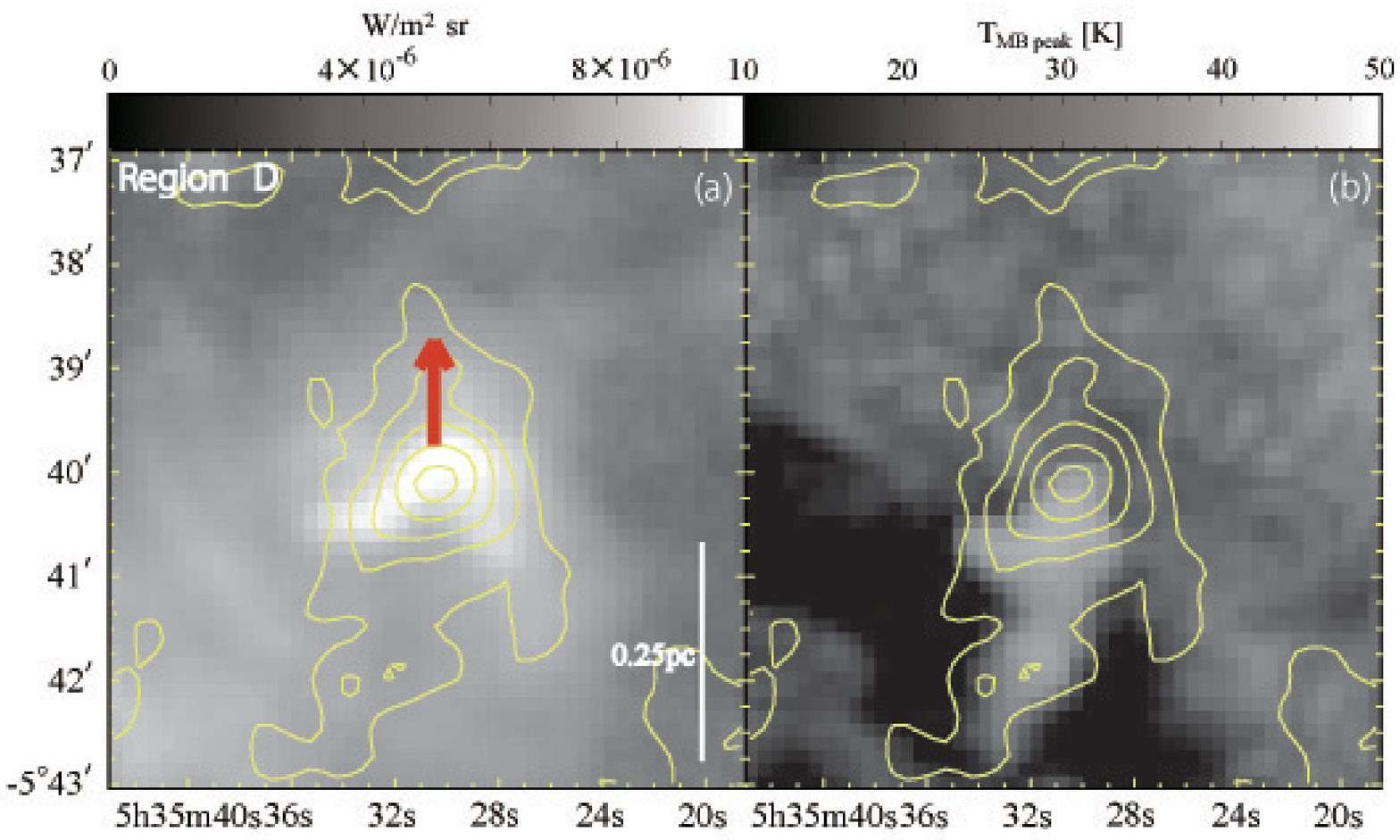} 
   \end{center}
 \footnotesize
   \caption{AzTEC/ASTE 1.1 mm contours overlied on a MSX 8 $\mu$m map and peak intensity map in the $^{12}$CO ($J$=1--0) line in Region D. 
Contour levels start from 10 $\sigma$ with an interval of 10 $\sigma$ for a range of 10 - 150 $\sigma$ and 50 $\sigma$ for a range from 150 $\sigma$.
}\label{aztec_8E}
\end{figure}

Figure \ref{aztec_8A} - \ref{aztec_8E} show the same maps as those in Figure \ref{pdr} but for Region A-D, respectively. 

In Region A where the 1.1 mm continuum peak locates at ($\alpha_{J2000}$, $\delta_{J2000}$)=($\timeform{5h34m 35s.5}, \timeform{-5D18'05".3}$), the nearest ($d$= $\sim$ 1.4 pc) HII region, M 42,  is located on the southeast from the continuum peak (see an arrow in Figure \ref{aztec_8A}).
 The 8 $\mu$m, 1.1 mm, and the $^{12}$CO emissions show the systematic distributions along the direction to the HII region;
The 8 $\mu$m emission distributes on the southeastern part of the 1.1 mm dust-continuum condensation, while the $^{12}$CO emission distributes on the central part of the 1.1 mm dust-continuum condensation. 
 
In Region B where the 1.1 mm continuum peak locates at ($\alpha_{J2000}$, $\delta_{J2000}$)=($\timeform{5h33m 57s.4}, \timeform{-5D23'28".9}$), the nearest ($d$= $\sim$ 2.3 pc) HII region, M 42,  is located on the east from the continuum peak (see an arrow in Figure \ref{aztec_8B}).
 The 8 $\mu$m and the $^{12}$CO emissions show the positional offset along the direction to the HII region, 
 and the 8 $\mu$m emission locates on the side of the HII region. 
The dusty condensation in Region B is associated with a Spitzer 24 $\mu$m source.

In Region C where the 1.1 mm continuum peak locates at ($\alpha_{J2000}$, $\delta_{J2000}$)=($\timeform{5h33m49s.9}, \timeform{-5D39'00".1}$), the nearest ($d$= $\sim$ 3.2 pc) HII region, M 42,  is located on the north from the continuum peak (see an arrow in Figure \ref{aztec_8C}).
 The 8 $\mu$m, $^{12}$CO, and the 1.1 mm emissions show the systematic distributions along the direction to the HII region; 
The 8 $\mu$m emission distributes on the northern part of the 1.1 mm dust-continuum condensation, while the $^{12}$CO emission  distribution is similar to the 1.1 mm emission distribution.

In Region D where the 1.1 mm continuum peak locates at ($\alpha_{J2000}$, $\delta_{J2000}$)=($\timeform{5h35m30s.5}, \timeform{-5D40'06".5}$), the nearest ($d$= $\sim$ 2.0 pc) HII region, M 42,  is located on the north from the continuum peak (see an arrow in Figure \ref{aztec_8E}).
 The 8 $\mu$m, 1.1 mm, and the $^{12}$CO emissions show the systematic distributions along the direction to the HII region; 
The 8 $\mu$m emission distributes on the northern part of the 1.1 mm dust-continuum condensation, while the $^{12}$CO emission distributes on the southern part of the 1.1 mm dust-continuum condensation.

\section{Discussion}

\subsection{Large Scale Possible External Effect; Collision of the Diffuse Components on the Cloud Surface}

\begin{figure}
   \begin{center}
      \FigureFile(100mm,100mm){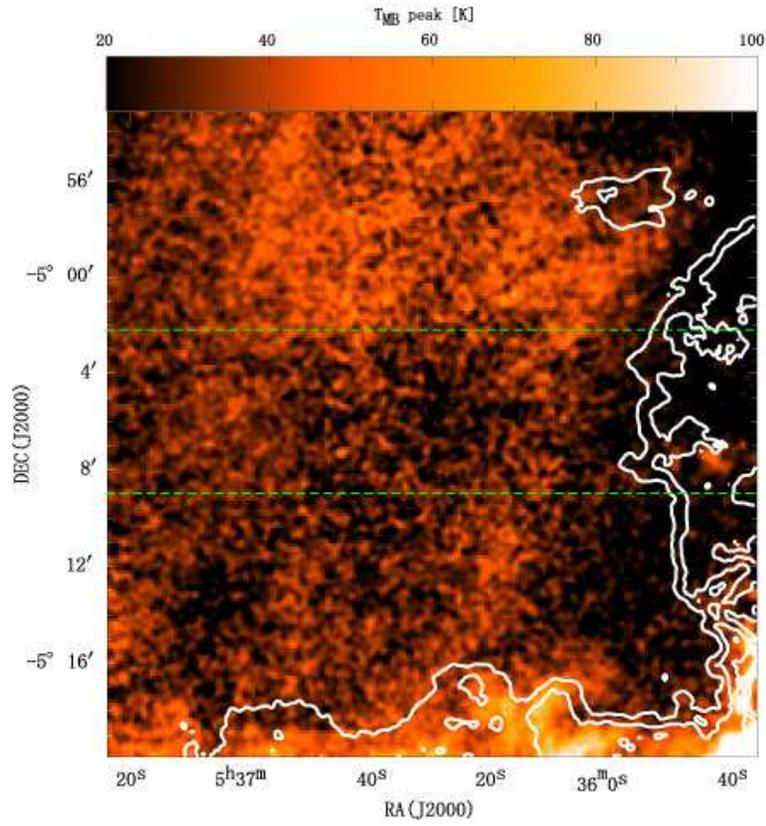} 
   \end{center}
 \footnotesize
   \caption{$^{12}$CO ($J$=1--0) peak intensity map (contour) superposed on the image of diffuse $^{12}$CO ($J$=1--0) components (color) in the east of the OMC-2/3 region.
A velocity range of the diffuse CO components is 2.6 - 8.6 km s$^{-1}$. 
The integrated velocity range of the image of diffuse CO components is 2.6 - 8.6 km s$^{-1}$. 
Contour levels are 15, 30, 45 K in $T_{\rm MB}$. 
}\label{diffuse}
\end{figure}

\begin{figure}
   \begin{center}
      \FigureFile(160mm,100mm){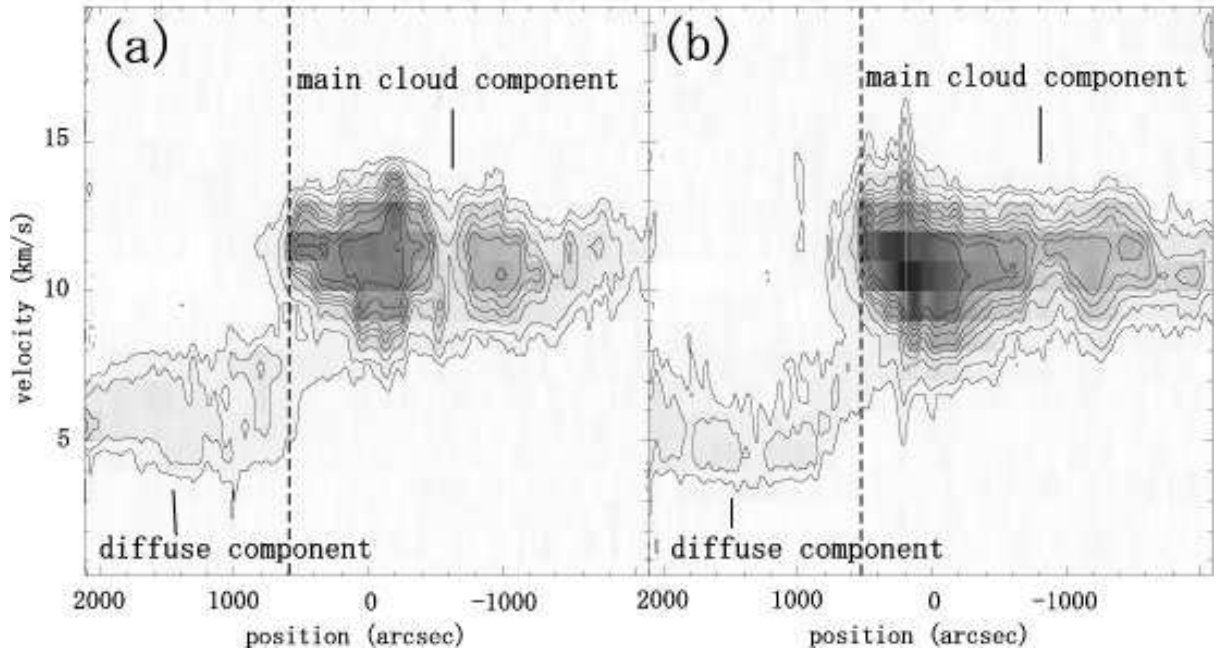}
   \end{center}
 \footnotesize
   \caption{Position-Velocity (P-V) diagrams of the $^{12}$CO ($J$=1--0) emission. The cut lines are shown in Figure \ref{diffuse}. The declination of the cut lines is 
$\timeform{-5D2'26".14}$ and $\timeform{-5D9'6".03}$, respectively. Vertical lines show the position of CO front. Contour levels start at 2 $\sigma$ levels with an interval of 2 $\sigma$.
}\label{RA_PV}
\end{figure}

\begin{figure}
   \begin{center}
      \FigureFile(160mm,100mm){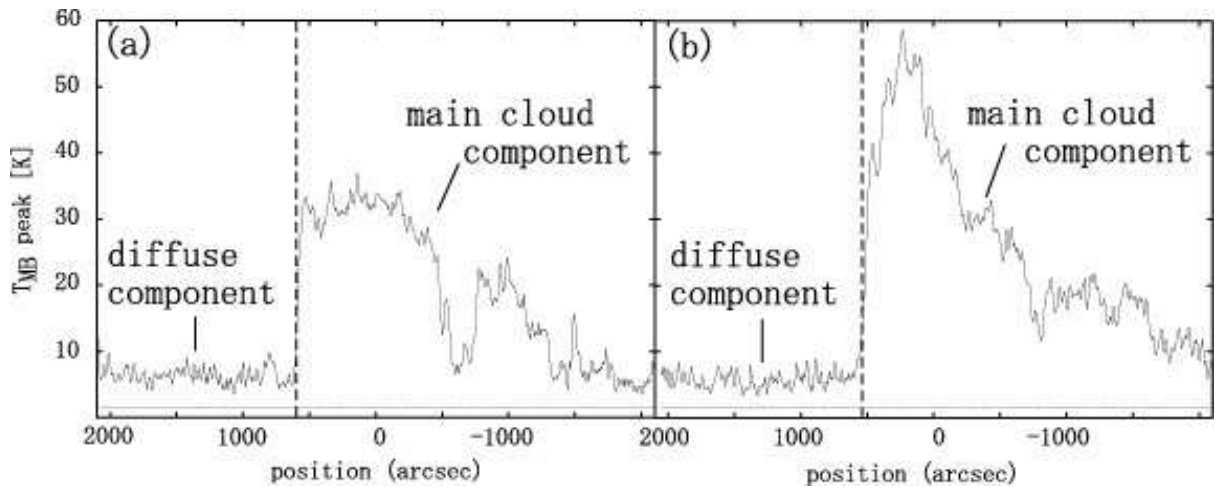}
   \end{center}
 \footnotesize
   \caption{Peak intensity profiles of the $^{12}$CO ($J$=1--0) emission. 
The cut lines is shown in Figure \ref{diffuse}. The declination of the cut lines are 
$\timeform{-5D2'26".14}$ and $\timeform{-5D9'6".03}$, respectively.  Vertical dash lines show the position of CO front. Horizontal lines show the noise level (1 $\sigma$ = 1.4 K at $T_{\rm MB}$) at the velocity channel maps with a velocity width of 0.4 km s$^{-1}$.    
}\label{int}
\end{figure}

One of the first findings in our extensive $^{12}$CO ($J$=1--0) emission observations of the northern part of the Orion A molecular cloud is the presence of the diffuse CO component over the entire eastern region of the integral-shape filament. 
Figure \ref{diffuse} highlights the total integrated intensity map of the diffuse CO component 
with respect to the peak intensity map at the OMC-2/3 region, and 
Figure \ref{RA_PV} shows position - velocity (P-V) diagrams whose cut lines are along the RA direction shown in Figure \ref{diffuse}. 
In the P-V diagrams, the central velocity of the diffuse CO component  is 5 - 6 km s$^{-1}$, 
while that of the main component is 11 - 12 km s$^{-1}$. 
These two velocity components appear to be connected at CO front, raising the $^{12}$CO velocity dispersion at CO front.  
Figure \ref{int} shows peak intensity profiles whose cut lines are the same as that of the P-V diagrams in Figure \ref{RA_PV}. 
While the peak intensity of the diffuse CO component is around 4 - 7 K at $T_{\rm MB}$, that of the main cloud component exceeds 50 K, and hence there is an abrupt increase of the $^{12}$CO peak intensity at CO front. 
In contrast, toward the western outskirts of the main cloud component the $^{12}$CO peak intensity decreases gradually (see Figure \ref{co_alla} - \ref{co_allc}). 
These peculiar features extend over the entire CO front as can be seen in Figures \ref{co_alla} - \ref{co_channel}, although toward the southern part of CO front the difference of the central velocities becomes smaller. 
The reason why the velocity difference between the diffuse CO component and the main cloud component becomes less clear is the large scale velocity gradient of the main cloud component (see \S 3.2).

There are two possible interpretations on the origin of the diffuse CO component; 
One is the trail of the main cloud component moving away from the galactic plane 
and the other is the pre-existing surrounding diffuse gas swept up by neighboring external sources.
\citet{Sakamoto97} has suggested these two possibilities on the origin of the diffuse CO component in the east of the L 1641 N region. 
It is, however, difficult to explain the difference of the central velocity between the diffuse CO component and the main cloud component and 
 the abrupt increase of the $^{12}$CO peak intensity by the former interpretation. 
We consider that the ``L-shaped'' feature of the diffuse CO component in the P-V diagrams and the abrupt increase of the $^{12}$CO peak intensity imply presence of the interaction between the diffuse CO component and the main cloud component \citep{Takakuwa03, Shimajiri08}, 
and that the diffuse CO component is being accumulated on to the surface of the main cloud component at CO front. 
In fact, \citet{Bally87}, \citet{Lee09} and \citet{Wilson05} have suggested that Ori OB 1a and Ori OB 1b, $\sim$ 100 pc away from the Orion A molecular cloud, affect the Orion A molecular cloud and disturb the morphology and kinematics of the cloud, respectively. 
These OB stars are the possible sources to sweep up the pre-existing diffuse material toward the direction to the main cloud component.
Our observational results favor the proposal by \citet{Wilson05}, since Ori OB 1b locates in the east of the Orion A molecular cloud and the diffuse CO component also distributes toward the same direction from the Orion A molecular cloud.
The collision of the diffuse CO component with the main cloud component might trigger the next star formation in the cloud \citep{Lada87}.

\subsection{Intermediate Scale Possible External Effect; Photo-Dominated Regions}

\begin{figure}
   \begin{center}
         \FigureFile(100mm,100mm){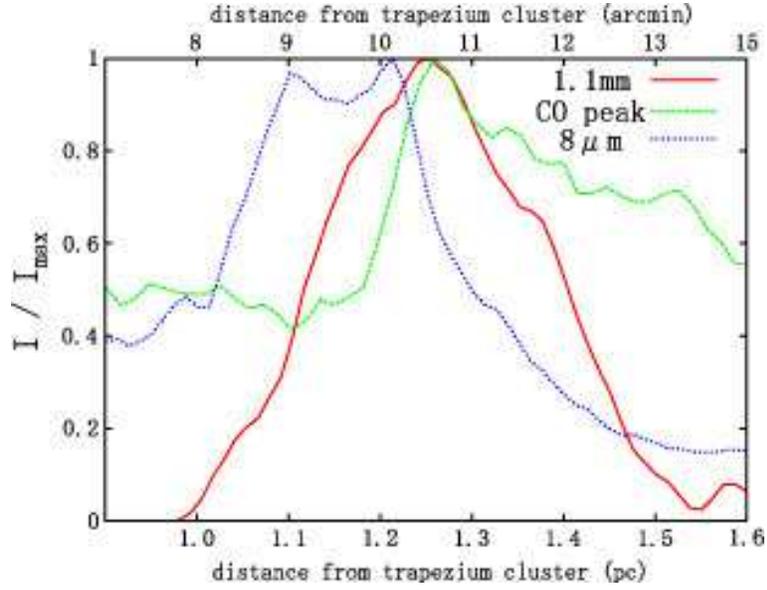} 
   \end{center}
 \footnotesize
   \caption{Intensity profiles of the 1.1 mm continuum emission, $^{12}$CO $T_{\rm MB}^{\rm peak}$ peak, and 8 $\mu$m emission, whose cut line is shown in Figure \ref{pdr}, as a function of the distance from trapezium cluster ($\theta^{1}$ Ori C).
}\label{int_profile}
\end{figure}

In the last section, we suggest the possibility that the diffuse CO component is swept up by Ori OB 1b and interacts with the entire Orion A molecular cloud. In this section, we focus on more local effects in a few pc-scale, namely, the HII regions embedded in the Orion A molecular cloud. 
As well as the external sources away from the cloud, newly born OB stars embedded in the cloud can also influence 
on the structure, chemistry, thermal balance, and the star formation in the cloud \citep{Hollenbach97}.
There are several OB stars (such as $\theta$$^1$ Ori C and NU Ori) and HII regions in our mapping area of the Orion A molecular cloud 
and these sources are likely to affect star formation in the Orion A molecular cloud. 
As shown in \S 3.3, we found that there are systematic differences of the distribution among the $^{12}$CO, 1.1 mm, and the 8 $\mu$m emissions and that the 8 $\mu$m, 1.1 mm, and the $^{12}$CO emissions are distributed as a function of the distance from the OB stars. 
The most clear example is in DLSF irradiated by $\theta$$^1$ Ori C, and the intensity profiles as a function of the distance from $\theta$$^1$ Ori C are shown in Figure \ref{int_profile}.

The presence of the strong 8 $\mu$m emission suggests the FUV radiation from the HII regions. 
On the other hand, $^{12}$CO molecules are dissociated by the FUV radiation, since the energy of FUV (6 eV - 13.6  eV) is high enough to dissociate $^{12}$CO molecules with an dissociative energy of 11.4 eV. 
Hence, the $^{12}$CO emission does not distribute at the near side of the exciting star. 
The 1.1 mm emission traces both the cold and warm dusts. 
Hence, the systematic distributions of the 8 $\mu$m, 1.1 mm, and the $^{12}$CO  emissions can be explained by the PDR model \citep{Hollenbach97}. 
In fact, toward DLSF and M 43, where the systematic distributions of the $^{12}$CO, 1.1 mm, and the 8 $\mu$m emissions are seen,  the observations of the CN line, one of the excellent tracers for PDR, suggest the presence of PDR. 

Region A-D, relatively isolated dusty condensations, also exhibit the systematic distributions of the $^{12}$CO, 1.1 mm, and the 8 $\mu$m emissions along the direction to the exciting sources. 
These isolated condensations are likely to be pre-existing molecular cloud cores.
When an expanding HII region interacts with a pre-existing molecular cloud core, the strong UV radiation will compress the core. This compression could induce the gravitational collapse of the core \citep{Bertoldi89, Motoyama07}.
Our results suggest that the dusty condensations of Region A-D are affected by the UV radiation from the OB stars. 
Moreover, the dusty condensation of Region B is associated with the spitzer 24 $\mu$m sources, suggesting the presence of the star formation.

\section{Summary}

 We have carried out extensive mapping observations
of the northern part of the Orion A molecular cloud in the 1.1 mm dust-continuum emission
with the AzTEC camera equipped in the ASTE telescope and 
in the $^{12}$CO ($J$=1--0) emission 
with the BEARS receiver equipped in the NRO 45 m telescope. 
 The main results of our wide-field ($\timeform{1D.7} \times \timeform{2D.3}$) and high-sensitivity ($\sim$ 9 mJy beam$^{-1}$)
 observations are summarized as follows; 

\begin{enumerate}

\item{
We have found new features of the 1.1 mm dust-continuum emission in addition to the previously known integral-shape filament. 
In the OMC-2/3 region, a new dusty filamentary structure is detected toward the east of the integral-shape filament. 
A dusty shell-like structure around the HII region, M 43, and a filamentary structure associated with DLSF are detected. 
In the southmost region where an active cluster-forming region of L 1641 N is located, 
four filamentary structures with a size of $\sim$ 0.5 - 2.0 pc are seen aligned almost in parallel.
}

\item{
In the $^{12}$CO ($J$=1--0) line observations, we found diffuse CO component over the entire eastern region of the integral-shape filament. 
The peak intensity profiles from the diffuse CO component to the integrals-shape filament show an abrupt increase at the border line (CO front) and the P-V diagrams show the sudden velocity change from 5 - 6  km s$^{-1}$ to 11- 12 km s$^{-1}$. 
We suggest that these peculiar features of the diffuse CO component show that the diffuse material swept up by Orion OB 1b, $\sim$100 pc away from the  Orion A molecular cloud,  is being  accumulated onto the surface of the integral-shape filament. 
}

\item{Comparisons among the $^{12}$CO, 1.1 mm, and the 8 $\mu$m emission distribution in M 43, DLSF, and Region A-D have revealed that the  8 $\mu$m, 1.1 mm, and the $^{12}$CO emissions are located 
sequentially as a function of the distance from the neighboring OB stars. 
These results suggest the presence of the PDRs around M 43, DLSF, and Region A-D, excited by the NU Ori, $\theta$$^1$ Ori C, and  $\theta$$^1$ Ori C, respectively. 
In particular, the dusty condensation of Region B is associated with a Spitzer 24 $\mu$m source, suggesting the presence of the star formation. 
Region A-D are possible site of star formation triggered by the PDRs.
}


\end{enumerate}

We thank the referee, Charles Lada, for his constructive comments that polish the manuscript significantly. 
We are grateful to the staffs at the Nobeyama Radio Observatory (NRO) for both operating the
45 m and helping us with the data reduction, 
and to NRO is a branch of the National Astronomical Observatory, National Institutes of Natural Sciences, Japan. We also acknowledge the ASTE staffs for both operating ASTE and helping us with the data reduction. 
Observations with ASTE were (in part) carried out remotely from Japan by using NTT's GEMnet2 and its partner R\&E (Research and Education) networks, which are based on AccessNova collaboration of University of Chile, NTT Laboratories, and National Astronomical Observatory of Japan. 
This research has made use of the NASA/ IPAC Infrared Science Archive, which is operated by the Jet Propulsion Laboratory, California Institute of Technology, under contract with the National Aeronautics and Space Administration.
This work was supported by Grant-in-Aid
for Scientific Research A 18204017.
Y. Shimajiri is financially supported by a Research Fellowship from the JSPS for Young Scientists.

\appendix
\section{AzTEC Data Reduction \& Performance}

In this appendix, we present our imaging simulation with artifact gaussian sources as input models, in order to test the performance of the AzTEC data reduction. 

\subsection{AzTEC Data Reduction}

The AzTEC data set is reduced by using the available AzTEC data reduction pipeline written in IDL and developed by the University of Massachusetts according to guidelines in \citet{Scott08}.
This pipeline includes the despiking, atmosphere removal, and optimal filtering. The despiking which is removal of the cosmic ray is,  however, not applied to 
Orion data reduction to avoid the removal of source emission, since the emission in the Orion region is too strong. 
Principal components analysis (PCA) cleaning method is applied for removal of the atmospheric noise in this pipeline. 
Finally, the 30 individual maps were coadded.
This data reduction pipeline has, however, been optimized for the identification of point sources. 
The PCA cleaning removes the astronomical signals in the case that the emission has an extended structure, 
since the signals which are a good correlation among each element are regarded as the atmospheric components and are removed. Hence, the flux and size of the extended sources are underestimated. 
In order to recover the extended features in the Orion map which is subtracted by the PCA cleaning, 
we apply an iterative mapping method to the Orion data-set. Hereafter, we call this method, FRUIT.
This method is based on an iterative mapping method applied to the BOLOCAM data reduction pipeline \citep{Enoch06}. 
The FRUIT data reduction pipeline is also written in IDL and developed by the University of Massachusetts. 
The FRUIT algorithm is described in \citet{Liu10}.

\subsection{Performance of FRUIT}

\begin{figure}
   \begin{center}
         \FigureFile(80mm,100mm){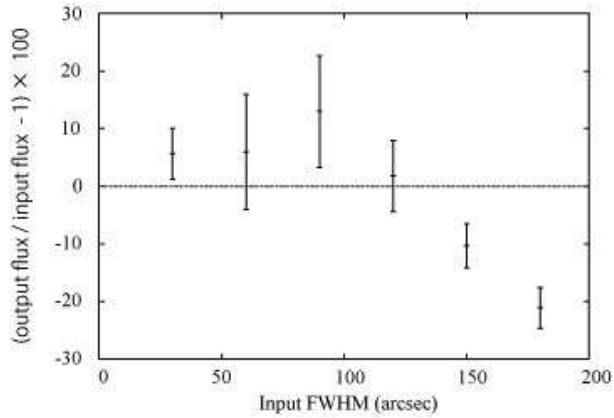} 
   \end{center}
 \footnotesize
   \caption{Fraction of the over-estimated (positive) or missed (negative)
flux as a function of the FWHM size of the input model Gaussian.   
}\label{recover}
\end{figure}

\begin{figure}
   \begin{center}
         \FigureFile(80mm,100mm){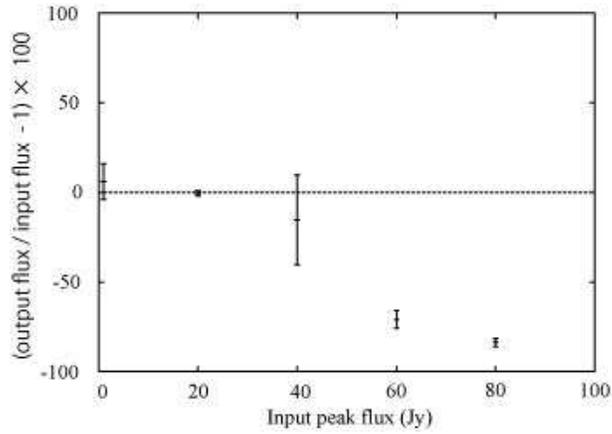} 
   \end{center}
 \footnotesize
   \caption{Fraction of the over-estimated (positive) or missed (negative)
flux as a function of the peak flux of the input model Gaussian.   
}\label{recover_strong}
\end{figure}

\begin{figure}
   \begin{center}
         \FigureFile(160mm,100mm){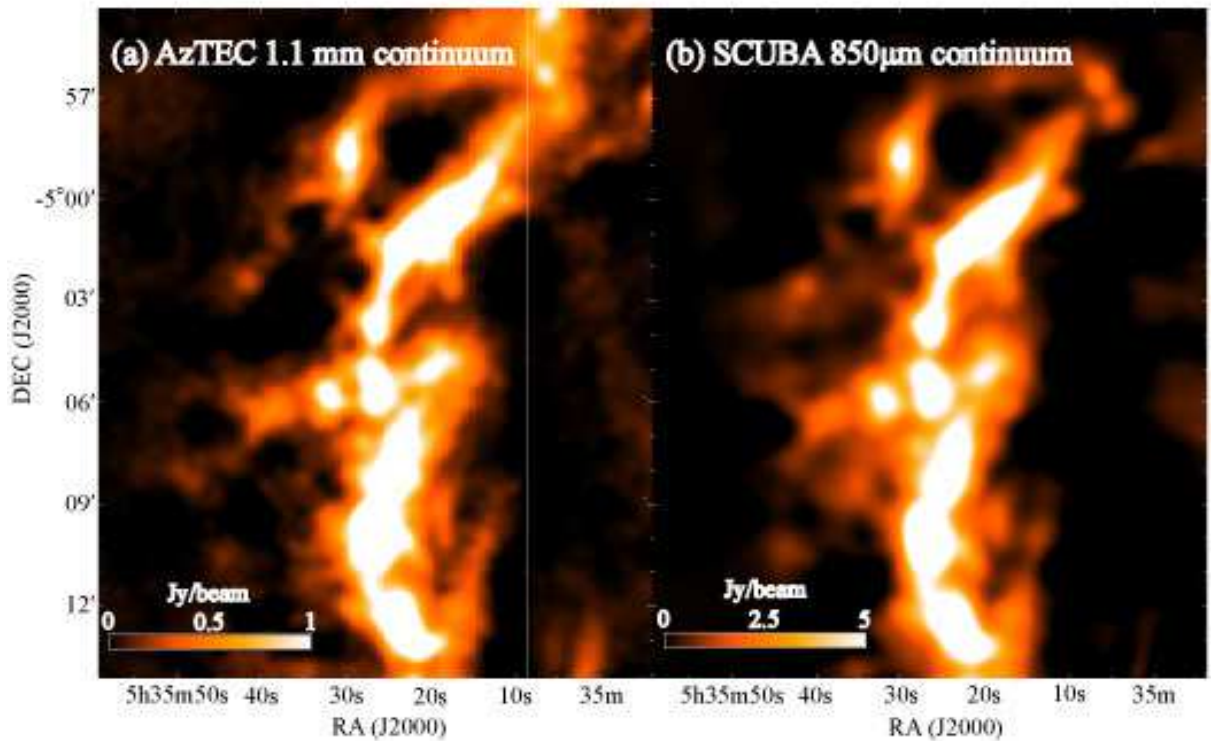}          
   \end{center}
 \footnotesize
   \caption{(a) AzTEC FRUIT map in the OMC-2/3 region. (b) SCUBA 850 $\mu$m map in the OMC-2/3 region obtained by \citet{Johnstone99}. 
}\label{aztec_scuba2}
\end{figure}

In order to investigate the performance of FRUIT, we have performed the simulation that the twenty five gaussian sources with a peak flux of 1 Jy and a FWHM size of 30, 60, 90, 120, 150, and 180$\arcsec$ are inserted to Orion data-set and  are applied the data reduction pipeline.  
Figure \ref{recover} shows the fraction of the over-estimated (positive) or missed (negative)
flux, 
$
(\frac{\mbox{output total flux}}{\mbox{input total flux}} - 1) \times 100, 
$
as a function of the FWHM size of the input model Gaussian. 
We have applied the aperture photometry to estimate the total flux. Error bars show a standard deviation (1 $\sigma$) of the output total flux among twenty five sources. 
In the case that the FWHM size of the input model Gaussian is under 150$\arcsec$, the output total flux is underestimated by less than 20\%. 
The total-flux reproductivity of the source with larger FWHM is lower than that with smaller FWHM. 
We insert the twenty five sources with a peak flux of 1, 20, 40, 60, and 80 Jy and a FWHM size of 60$\arcsec$ to the Orion data-set and applies the same data reduction pipeline. 
Figure \ref{recover_strong} shows the fraction of the over-estimated (positive) or missed (negative)
flux as a function of the peak flux of the input model Gaussian.
The total flux of sources with a peak flux under 20 Jy is underestimated by less than 10 \%. 
The reproductivity of the total flux is, however, low for the source with a peak flux over 20 Jy.
In the case of Orion, the total flux of sources in our map should be recovered by more than 80 \%, since the peak flux and the FWHM size of most sources are smaller than 20 Jy and 150$\arcsec$. 
Moreover, the restored image of the 1.1 mm dust continuum emission is consistent with that of the SCUBA 850 $\mu$m dust continuum emission as shown in Figure \ref{aztec_scuba2}.






\clearpage

\begin{figure}
   \begin{center}
          \FigureFile(80mm,100mm){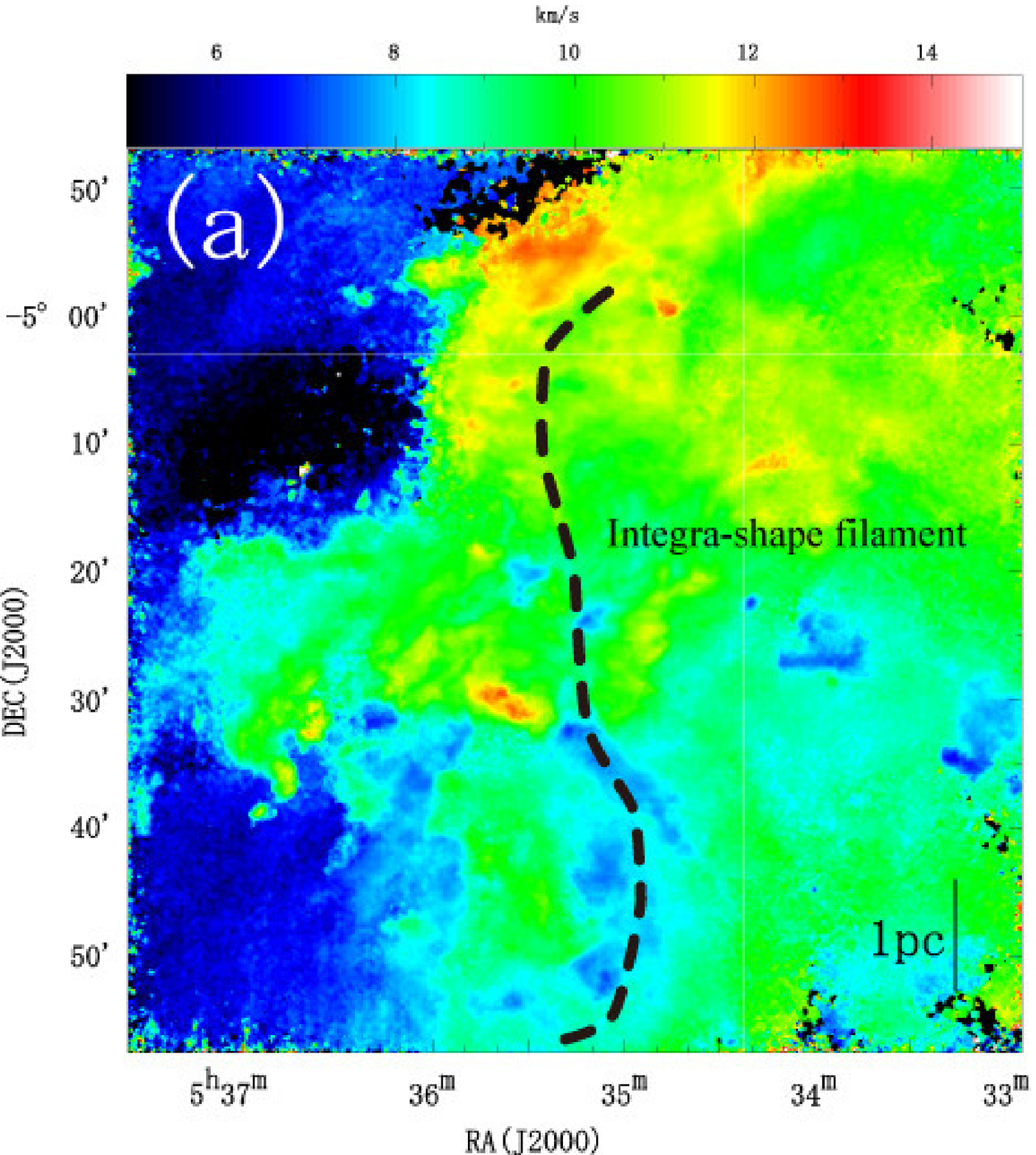}
              \FigureFile(80mm,100mm){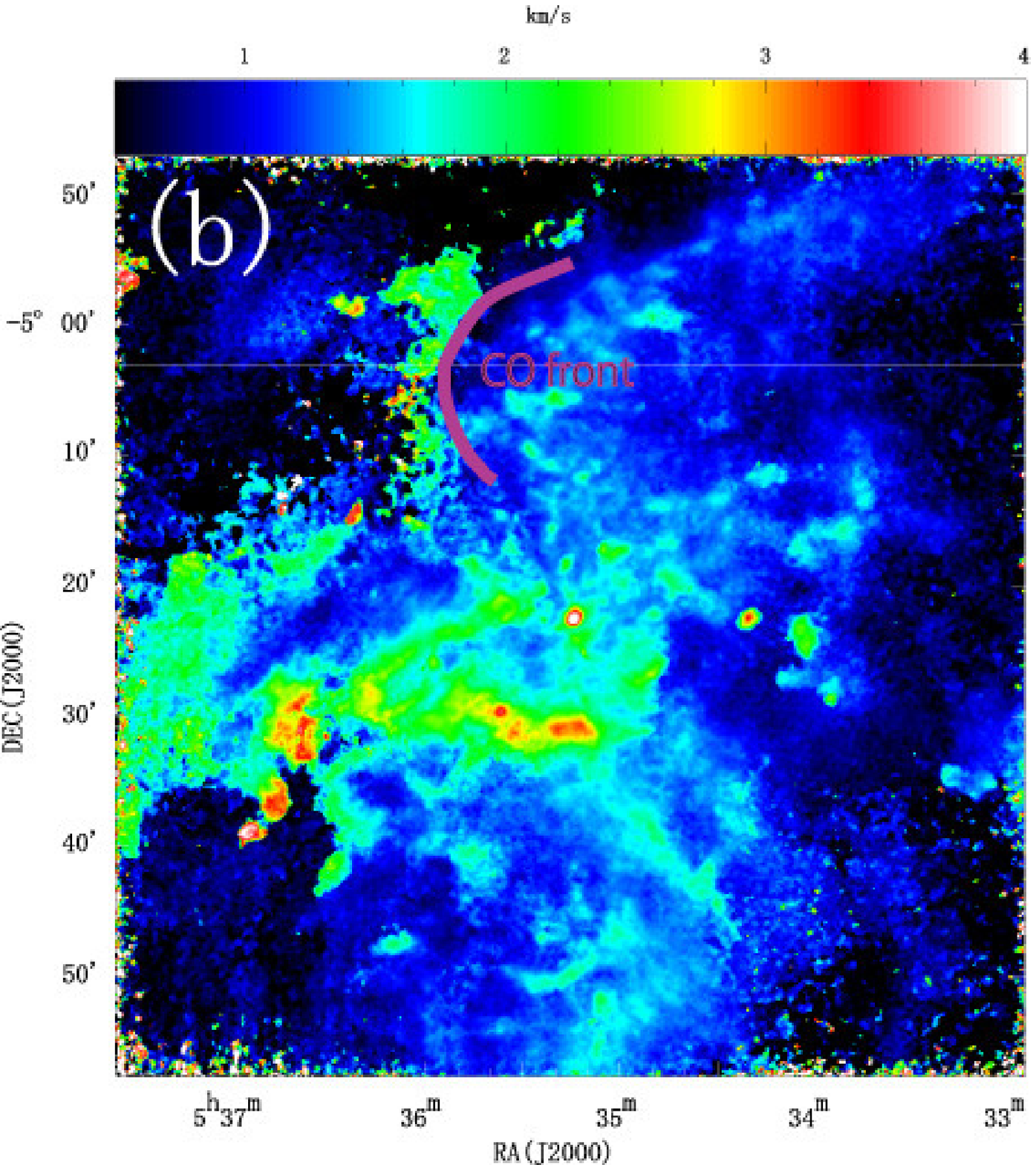}
   \end{center}
 \footnotesize
   \caption{(a) $^{12}$CO intensity-weighted mean velocity map at a velocity range of 0.0 - 20.0 km s$^{-1}$. Blue, green, and yellow colors show the $\le$ 9.0, 9.0 - 11.0 , $\ge$ 11.0 km s$^{-1}$, respectively. (b) $^{12}$CO velocity dispersion map at a velocity range of 0.0 - 20.0 km s$^{-1}$. Blue, green, and yellow colors shows the $\le$ 1.6, 1.6 - 2.8, $\ge$ 2.8 km s$^{-1}$, respectively. 
}\label{co_all2}
\end{figure}

 \clearpage

 \begin{table}
 \begin{center}
\caption{PARAMETERS FOR OBSERVATIONS}
 \label{obs}
 \begin{tabular}{lcc}
 \hline
  Telecope/Receiver & ASTE / AzTEC & NRO 45 m / BEARS \\
  \hline
  Line / Wavelength & 1.1 mm & $^{12}$CO ($J$=1--0;115.271 GHz) \\
  Observation date & Oct. 2008 - Dec. 2008 & Nov. 2007 - May 2008 \\
  Observing mode & raster & OTF \\
  Observation time & 15 hr & 40 hr \\
  Mapping size &  $\timeform{1.D7}$ $\times$ $\timeform{2D.3}$  & $\timeform{1.D2}$ $\times$ $\timeform{1D.2}$ \\
  Effective beam size & 40$\arcsec$ & 21$\arcsec$ \\
  Velocity width & ------ & 1.0  km s$^{-1}$ \\
  Typical r.m.s. & 9 mJy beam$^{-1}$ & 0.9 K \\
 \hline
\multicolumn{3}{@{}l@{}}{\hbox to 0pt{\parbox{180mm}{\footnotesize
}\hss}}
  \end{tabular}
\end{center}
\end{table}


\begin{thebibliography}{}
\bibitem[Aso et al.(2000)]{Aso00} Aso, Y., Tatematsu, K., Sekimoto, Y., Nakano, T., Umemoto, T., Koyama, K., \& Yamamoto, S.\ 2000, \apjs, 131, 465 

\bibitem[Bally et al.(1987)]{Bally87} Bally, J., Lanber, W.~D., Stark, A.~A., \& Wilson, R.~W.\ 1987, \apjl, 312, L45 

\bibitem[Bertoldi(1989)]{Bertoldi89} Bertoldi, F.\ 1989, \apj, 346, 735 

\bibitem[Bernstein et al.(1999)]{Bernstein99} Bernstein, M.~P., Sandford, S.~A., Allamandola, L.~J., Gillette, J.~S.~B., Clemett, S.~J.,\& Zare, R.~N.\ 1999, Science, 283, 1135 

\bibitem[Chen et al.(1995)]{Chen95} Chen, H., Zhao, J.-H., \& Ohashi, N.\ 1995, \apjl, 450, L71 

\bibitem[Chen et al.(1996)]{Chen96} Chen, H., Ohashi, N., \& Umemoto, T.\ 1996, \aj, 112, 717 


\bibitem[Chini et al.(1997)]{Chini97} Chini, R., Reipurth, B., Ward-Thompson, D., Bally, J., Nyman, L.-A., Sievers, A., \& Billawala, Y.\ 1997, \apjl, 474, L135 

\bibitem[Cowie et al.(1979)]{Cowie79} Cowie, L.~L., Songaila, A., \& York, D.~G.\ 1979, \apj, 230, 469 

\bibitem[Davis et al.(2009)]{Davis09} Davis, C.~J., et al.\ 2009, \aap, 496, 153 

\bibitem[Elmegreen(1998)]{Elmegreen98} Elmegreen, B.~G.\ 1998, Origins, 148, 150 

\bibitem[Enoch et al.(2006)]{Enoch06} Enoch, M.~L., et al.\ 2006, \apj, 638, 293 

\bibitem[Ezawa et al.(2004)]{Ezawa04} Ezawa, H., Kawabe, R., Kohno, K., \& Yamamoto, S.\ 2004, \procspie, 5489, 763 

\bibitem[Goudis et al. (1982)]{Goudis82} Goudis: 1982, $The Orion complex : a case study of interstellar matter, D. Reidel Publishing Company$ 

\bibitem[Heyer et al.(1992)]{Heyer92} Heyer, M.~H., Morgan, J., Schloerb, F.~P., Snell, R.~L., \& Goldsmith, P.~F.\ 1992, \apjl, 395, L99 

\bibitem[Hirota et al.(2008)]{Hirota08} Hirota, T., et al.\ 2008, \pasj, 60, 961 

\bibitem[Hollenbach \& Tielens(1997)]{Hollenbach97} Hollenbach, D.~J., \& Tielens, A.~G.~G.~M.\ 1997, \araa, 35, 179 

\bibitem[Homeier \& Alves(2005)]{Homeir05} Homeier, N.~L., \& Alves, J.\ 2005, \aap, 430, 481 

\bibitem[Hosokawa \& Inutsuka(2005)]{Hosokawa05} Hosokawa, T., \& Inutsuka, S.-i.\ 2005, \apj, 623, 917 

\bibitem[Ikeda et al.(2007)]{Ikeda07} Ikeda, N., Sunada,K., \& Kitamura, Y.\ 2007, \apj, 665, 1194 

\bibitem[Johnstone \& Bally(1999)]{Johnstone99} Johnstone, D., \& Bally, J.\ 1999, \apjl, 510, L49 

\bibitem[Kohno et al.(2004)]{Kohno04} Kohno, K., et al.\ 2004, The Dense Interstellar Medium in Galaxies, 349 

\bibitem[Kutner \& Ulich(1981)]{Kutner81} Kutner, M.~L., \& Ulich, B.~L.\ 1981, \apj, 250, 341 

\bibitem[Lada(1987)]{Lada87} Lada, C.~J.\ 1987, Star Forming Regions, 115, 1 

\bibitem[Lada \& Lada(2003)]{Lada03} Lada, C.~J., \& Lada, E.~A.\ 2003, \araa, 41, 57 

\bibitem[Lee \& Chen(2009)]{Lee09} Lee, H.-T., \& Chen, W.~P.\ 2009, \apj, 694, 1423 

\bibitem[Lis et al.(1998)]{Lis98} Lis, D.~C., Serabyn, E., Keene, J., Dowell, C.~D., Benford, D.~J., Phillips, T.~G., Hunter, T.~R., \& Wang, N.\ 1998, \apj, 509, 299 

\bibitem[Liu et al.(2010)]{Liu10} Liu, G., et al.\ 2010, \aj, 139, 1190 

\bibitem[Megeath(1994)]{Megeath94} Megeath, S.~T.\ 1994, The Structure and Content of Molecular Clouds, 439, 215 

\bibitem[Menten et al.(2007)]{Menten07} Menten, K.~M., Reid, M.~J., Forbrich, J., \& Brunthaler, A.\ 2007, \aap, 474, 515 

\bibitem[Momose et al.(1998)]{Momose98} Momose, M., Ohashi, N., Kawabe, R., Nakano, T., \& Hayashi, M.\ 1998, \apj, 504, 314 

\bibitem[Motoyama et al.(2007)]{Motoyama07} Motoyama, K., Umemoto, T., \& Shang, H.\ 2007, \aap, 467, 657

\bibitem[Myers et al.(1995)]{Myers95} Myers, P.~C., Bachiller, R., Caselli, P., Fuller, G.~A., Mardones, D., Tafalla, M., \& Wilner, D.~J.\ 1995, \apjl, 449, L65 

\bibitem[Nakamura \& Li(2007)]{Nakamura07} Nakamura, F., \& Li, Z.-Y.\ 2007, \apj, 662, 395 


\bibitem[Nutter \& Ward-Thompson(2007)]{Nutter07} Nutter, D., \& Ward-Thompson, D.\ 2007, \mnras, 374, 1413 

\bibitem[Rodr{\'{\i}}guez-Franco et al.(2001)]{Rod01} Rodr{\'{\i}}guez-Franco, A., Wilson, T.~L., Mart{\'{\i}}n-Pintado, J., \& Fuente, A.\ 2001, \apj, 559, 985 

\bibitem[Sakamoto et al.(1997)]{Sakamoto97} Sakamoto, S., Hasegawa, T., Hayashi, M., Morino, J.-I., \& Sato, K.\ 1997, \apj, 481, 302 

\bibitem[Sandell \& Knee(2001)]{Sandell01} Sandell, G., \& Knee, L.~B.~G.\ 2001, \apjl, 546, L49 

\bibitem[Saito et al.(1999)]{Saito99} Saito, M., Sunada, K., Kawabe, R., Kitamura, Y., \& Hirano, N.\ 1999, \apj, 518, 334 

\bibitem[Sandstrom et al.(2007)]{Sandstrom07} Sandstrom, K.~M., Peek, J.~E.~G., Bower, G.~C., Bolatto, A.~D., \& Plambeck, R.~L.\ 2007, \apj, 667, 1161 

\bibitem[Sawada et al.(2008)]{Sawada08} Sawada, T., et al.\ 2008, \pasj, 60, 445 

\bibitem[Scott et al.(2008)]{Scott08} Scott, K.~S., et al.\ 2008, \mnras, 385, 2225 

\bibitem[Shimajiri et al.(2008)]{Shimajiri08} Shimajiri, Y., 
Takahashi, S., Takakuwa, S., Saito, M., \& Kawabe, R.\ 2008, \apj, 683, 255

\bibitem[Shimajiri et al.(2009)]{Shimajiri09} Shimajiri, Y., Takahashi, S., Takakuwa, S., Saito, M., \& Kawabe, R.\ 2009, \pasj, 61, 1055 


\bibitem[Solomon et al.(1977)]{Solomon77} Solomon, P.~M., Sanders, D.~B., \& Scoville, N.~Z.\ 1977, \baas, 9, 554 

\bibitem[Sorai et al.(2000)]{Sorai00} Sorai, K., Sunada, K., Okumura, S.~K., Tetsuro, I., Tanaka, A., Natori, K., \& Onuki, H.\ 2000, \procspie, 4015, 86 

\bibitem[Stanke et al.(2002)]{Stanke02} Stanke, T., McCaughrean, M.~J., \& Zinnecker, H.\ 2002, \aap, 392, 239 

\bibitem[Stanke \& Williams(2007)]{Stanke07} Stanke, T., \& Williams, J.~P.\ 2007, \aj, 133, 1307 

\bibitem[Sunada et al.(2000)]{Sunada00} Sunada, K., Yamaguchi, C., Nakai, N., Sorai, K., Okumura, S.~K., \& Ukita, N.\ 2000, \procspie, 4015, 237

\bibitem[Takahashi et al.(2008)]{Takahashi08} Takahashi, S., Saito, M., Ohashi, N., Kusakabe, N., Takakuwa, S., Shimajiri, Y., Tamura, M., \& Kawabe, R.\ 2008, \apj, 688, 344 

\bibitem[Takahashi et al.(2009)]{Takahashi09} Takahashi, S., Ho, P.~T.~P., Tang, Y.-W., Kawabe, R., \& Saito, M.\ 2009, \apj, 704, 1459 

\bibitem[Takakuwa et al.(2003)]{Takakuwa03} Takakuwa, S., Ohashi, N., \& Hirano, N.\ 2003, \apj, 590, 932 

\bibitem[Takakuwa et al.(2004)]{Takakuwa04} Takakuwa, S., et al.\ 2004, \apjl, 616, L15 

\bibitem[Tatematsu et al.(1993)]{Tatematsu93} Tatematsu, K., et 
al.\ 1993, \apj, 404, 643 

\bibitem[Tatematsu et al.(1998)]{Tatematsu98} Tatematsu, K., 
Umemoto, T., Heyer, M.~H., Hirano, N., Kameya, O., \& Jaffe, D.~T.\ 1998, \apjs, 118, 517

\bibitem[Tatematsu et al.(2008)]{Tatematsu08} Tatematsu, K., 
Kandori, R., Umemoto, T., \& Sekimoto, Y.\ 2008, \pasj, 60, 407 

\bibitem[Thum et al.(1978)]{Thum78} Thum, C., Lemke, D., Fahrbach, U., \& Frey, A.\ 1978, \aap, 65, 207 

\bibitem[Tsujimoto et al.(2002)]{Tsujimoto02} Tsujimoto, M., Koyama, K., Tsuboi, Y., Goto, M., \& Kobayashi, N.\ 2002, \apj, 566, 974

\bibitem[Whitworth et al.(1994)]{Whitworth94} Whitworth, A.~P., Bhattal, A.~S.,Chapman, S.~J., Disney, M.~J., \& Turner, J.~A.\ 1994, \aap, 290, 421

\bibitem[Wilson et al.(2005)]{Wilson05} Wilson, B.~A., Dame, T.~M., Masheder, M.~R.~W., \& Thaddeus, P.\ 2005, \aap, 430, 523 

\bibitem[Wilson et al.(2008)]{Wilson08} Wilson, G.~W., et al.\ 2008, \mnras, 386, 807 

\bibitem[Yamaguchi et al.(2000)]{Yamaguchi00} Yamaguchi, C., Sunada, K., Iizuka, Y., Iwashita, H., \& Noguchi, T.\ 2000, \procspie, 4015, 614 

\bibitem[Yokogawa et al.(2003)]{Yokogawa03} Yokogawa, S., Kitamura, Y., Momose, M., \& Kawabe, R.\ 2003, \apj, 595, 266 

\bibitem[Yonekura et al.(2005)]{Yonekura05} Yonekura, Y., Asayama, S., Kimura, K., Ogawa, H., Kanai, Y., Yamaguchi, N., Barnes, P.~J., \& Fukui, Y.\ 2005, \apj, 634, 476 

\bibitem[Zavagno et al.(2006)]{Zavagno06} Zavagno, A., Deharveng, L., Comer{\'o}n, F., Brand, J., Massi, F., Caplan, J., \& Russeil, D.\ 2006, \aap, 446, 171 

\bibitem[Zinnecker \& Yorke(2007)]{Zinnecker07} Zinnecker, H., \& Yorke, H.~W.\ 2007, \araa, 45, 481


\end{thebibliography}
\end{document}